\documentclass[twocolumn,superscriptaddress,prb]{revtex4-1}

\usepackage{amsmath}
\usepackage{amsfonts}
\usepackage{color}
\usepackage[colorlinks,bookmarks=false,citecolor=blue,linkcolor=red,urlcolor=blue]{hyperref}
\usepackage{ulem}
\usepackage{braket}
\setcitestyle{numbers,square}
\usepackage{overpic}
\usepackage{subfig}

\newcommand{\RNum}[1]{\uppercase\expandafter{\romannumeral #1\relax}}

\def \be {\begin{equation}}
\def \ee {\end{equation}}
\def \ba {\begin{array}}
\def \ea {\end{array}}
\def \bea {\begin{eqnarray}}
\def \eea {\end{eqnarray}}

\def \ble {\begin{widetext}\begin{equation}}
\def \ele {\end{equation}\end{widetext}}
\def \blea {\begin{widetext}\begin{eqnarray}}
\def \elea {\end{eqnarray}\end{widetext}}

\def \a {\alpha}

\def \g {\gamma}

\def \s {\sigma}

\def \f {\frac}

\def \lag {\langle}
\def \rag {\rangle}

\def \Tr {{\mathrm{Tr}}}

\def \and {{\mathrm{and}}}

\begin{document}


\title{Identifying quantum many-body integrability and chaos using eigenstates trace distances}

\author{Reyhaneh Khasseh}

\affiliation{Theoretical Physics III, Center for Electronic Correlations and Magnetism,
Institute of Physics, University of Augsburg, D-86135 Augsburg, Germany}
\author{Jiaju Zhang}

\affiliation{Center for Joint Quantum Studies and Department of Physics, School of Science,
Tianjin University, 135 Yaguan Road, Tianjin 300350, China}

\author{Markus Heyl}
\affiliation{Theoretical Physics III, Center for Electronic Correlations and Magnetism,
Institute of Physics, University of Augsburg, D-86135 Augsburg, Germany}

\author{M.~A.~Rajabpour}
\affiliation{Instituto de Fisica, Universidade Federal Fluminense,\\
Av.~Gal.~Milton Tavares de Souza s/n, Gragoat\'a, 24210-346, Niter\'oi, RJ, Brazil}

\begin{abstract}
While the concepts of quantum many-body integrability and chaos are of fundamental importance for the understanding of quantum matter, their precise definition has so far remained an open question.
In this work, we introduce an alternative indicator for quantum many-body integrability and chaos, which is based on the statistics of eigenstates by means of nearest-neighbor subsystem trace distances.
We show that this provides us with a faithful classification through extensive numerical simulations for a large variety of paradigmatic model systems including random matrix theories, free fermions, Bethe-ansatz solvable systems, and models of many-body localization.
While existing indicators, such as those obtained from level-spacing statistics, have already been utilized with great success, they also face limitations.
This concerns for instance the quantum many-body kicked top, which is exactly solvable but classified as chaotic in certain regimes based on the level-spacing statistics, while our introduced indicator signals the expected quantum many-body integrability.
%
%
We discuss the universal behaviors we observe for the nearest-neighbor trace distances and point out that our indicator might be useful also in other contexts such as for the many-body localization transition.
\end{abstract}

\maketitle


Quantum chaos and integrability have been a major focus of research for decades due to their key relevance for the foundations of statistical physics and fundamental concepts such as thermalization. In classical physics, chaos manifests as a divergence of initially close-by phase-space trajectories. Integrability as a counterpart of chaos is defined by the existence of a maximal number of Poisson-commuting invariants~\cite{Babelon:2003,Gutzwiller:2013}. However, establishing a precise measure of quantum chaos and integrability in the \textit{quantum many-body regime} has remained an outstanding challenge~\cite{Caux:2011}. The most widely used indicator is based on level spacing statistics~\cite{Tabor:1977,Bohigas:1984}. 
However, it predicts chaotic behavior for some systems, which are expected to be considered integrable in the many-body sense~\cite{Finkel:2005,Sieberer:2019}. 

In this letter, we introduce an alternative indicator of quantum integrability and many-body quantum chaos based on the eigenstate properties instead of the spectrum. We show that our indicator correctly classifies a wide range of systems as quantum integrable, including Bethe-ansatz solvable models, quantum spin chains in a fully many-body localized (MBL) regime, and quadratic fermionic systems. It is a central result of this work that our indicator detects quantum integrability also in cases where the level spacing fails, such as the quantum many-body kicked-top model. Our indicator is based on subsystem trace distances between nearest-neighboring Hamiltonian eigenstates, which provides a bound on the smoothness of operator expectation values as a function of energy and, therefore, a natural connection to the eigenstate thermalization hypothesis (ETH). This measure is much more robust to the symmetries of the systems as compared to the level spacing statistics. 
This can be useful when the symmetries of a model are not fully understood.

To investigate whether a general quantum Hamiltonian exhibits chaos or integrability, we evaluate the trace norm distance between two reduced density matrices defined as
\begin{equation}\label{eq:dis}
D^{A}_{n}=\frac{1}{2}||\rho^{A}_{n+1}-\rho^{A}_n||_1~.
\end{equation} 
Here, $\rho_n^A=\Tr_{\bar{A}}\rho_n$ denotes the reduced density matrix of a subsystem $A$ and $\rho_n=\ket{\psi_n}\bra{\psi_n}$ the density matrix of an eigenstate $\ket{\psi_n}$ of a given Hamiltonian. We order the eigenstates $\ket{\psi_n}$ with respect to their eigenvalues $\epsilon_n$ in ascending order, i.e., $\epsilon_{n+1}>\epsilon_n$. While distances between density matrices can be defined in various ways~\cite{Dodonov:2000,Gilchrist:2005,Nielsen:2009}, the definition in Eq.~(\ref{eq:dis}) turns out to be practically suitable, as we will discuss in the remainder of this letter. In particular, it has been found that $D_n^A$ provides a general upper bound on the smoothness of operator expectation values as a function of energy~\cite{Nielsen:2009,Rajabpour:2020}:

\begin{equation}\label{eq:dis_noneq}
|\Delta {\cal O}_n|=|\Tr(\rho^{A}_{n+1}-\rho^{A}_n){\cal O}|\leq 2sD^{A}_{n}.
\end{equation} 
Here, $\Delta {\cal O}_n=\bra{\psi_{n+1}}{\cal O}\ket{\psi_{n+1}}-\bra{\psi_{n}}{\cal O}\ket{\psi_{n}}$ denotes the difference of operator expectation values in neighboring eigenstates with ${\cal O}$ an operator defined in subsystem $A$ and $s$ is the largest singular value of the operator ${\cal O}$~\cite{Nielsen:2009}. The expectation values of local observables for various Hamiltonian eigenstates may fluctuate between neighboring eigenstates and conform to thermal predictions if ETH is valid. This distinction can be used to differentiate between integrable and chaotic models. The ETH, which pertains to chaotic Hamiltonians, stipulates that the diagonal matrix elements of observables within Hamiltonian eigenstates exhibit a smooth energy dependence and a narrow distribution. Conversely, in the integrable regime, the expectation values across the spectrum tend to fluctuate significantly, as previously demonstrated in various research studies (see references~\cite{Rigol:2008} and~\cite{Deutsch:2018} for a review). We propose to use the eigenstate trace distances measure in Eq.~(\ref{eq:dis}), which expands upon the definition of the ETH based on subsystem trace distances in Ref.~\cite{Dymarsky:2016ntg} by providing quantitative criteria for quantum many-body chaos and integrability. 
%


In the following, we will introduce the microscopic chaotic and integrable models we use to illustrate our findings.

{\it{Many-body quantum chaotic systems}}: Let us start with the analysis of many-body quantum chaotic systems.
In this context, we will use for our analysis the paradigmatic quantum Ising chain with both transverse and longitudinal fields
\begin{equation}\label{eq:Hamiltonian_Ising}
H_{Ising}=\sum_{l=1}^{L}(J\sigma_l^z\sigma_{l+1}^z+h_z\sigma_l^z+h_x\sigma_l^x)~,
\end{equation}
where $\sigma_l^{\alpha}(\alpha=x,y,z)$ denote the Pauli spin operators at site $l$, $J$ is the coupling constant, and $h_{\beta}(\beta=x,z)$ represent the strengths of the two magnetic fields. In what follows, we will set  the interaction $J = 1$. We assume periodic boundary conditions (PBC), i.e.,  $\sigma^{\alpha}_{L+1}=\sigma^{\alpha}_{1}(\alpha=x,y,z)$, which implies translational invariance. This symmetry enables us to partition the Hamiltonian into different sectors with a conserved momentum of $K=2\pi j/L$, where $j=0,..., L-1$. Each sector can be independently diagonalized, reducing the computational complexity.  It is worth noting that the main features of the statistics of $D_{n}^A$ are identical for single symmetry sectors or for the full spectrum. We, therefore, focus on a single sector $K=2\pi/L$ without loss of generality. Let us mention, however, that the $K=0,\pi$ sectors provide an exception due to the presence of further symmetries. Examples of different symmetry blocks and comparisons with results from the full spectrum can be found in the Supplemental Material~\cite{Supplement}. In the following, it will be suitable to analyze $D_n^A$ as a function of $x=L_A/L$, so that we introduce the following notation:
\begin{equation}
    D_{n}(x)=\frac{1}{2}||\rho_{n+1}(x)-\rho_n(x)||_1, \quad x=L_A/L \, ,
\end{equation}
in addition to Eq.~(\ref{eq:dis}).
\begin{figure}[htp]\label{fig:statistics_D_Ising} 
\includegraphics[width=80mm]{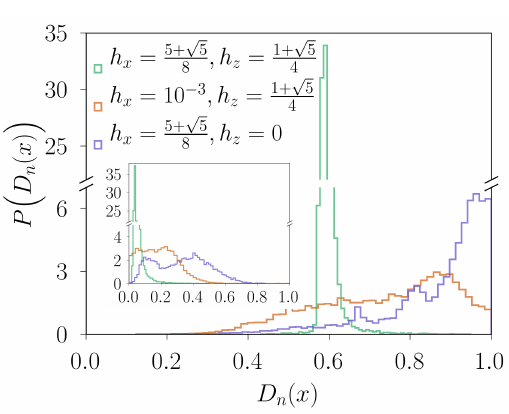}\put(-150,100){$x=4/18$}\put(-60,150){$x=1/2$}
	\caption{{\bf Ising model}: Distribution $P(D_{n}(x))$ of trace distances $D_{n}(x)$ for nearest-neighbor eigenstates with $x=L_A/L$ the ratio between subsystem size $L_A$ and system size $L=18$. The main plot shows data for $P(D_{n}(x))$ at different magnetic field strengths for $L_A=L/2$ and the inset for $L_A=4$, focusing on one symmetry sector: $K = 2\pi/L$.}
	\label{fig:dis_ave_comp}
\end{figure}
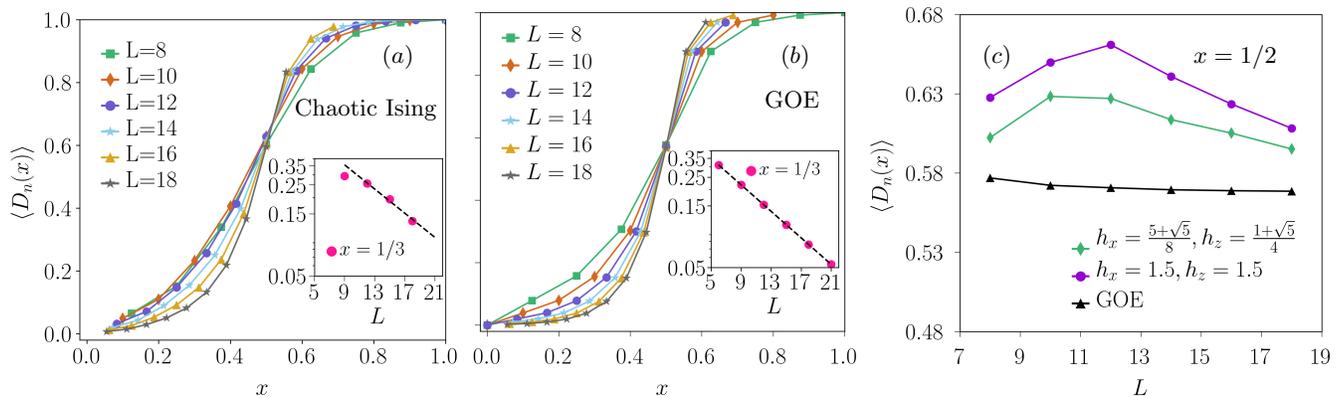
\begin{figure*}[htp]
\includegraphics[width=61.5mm]{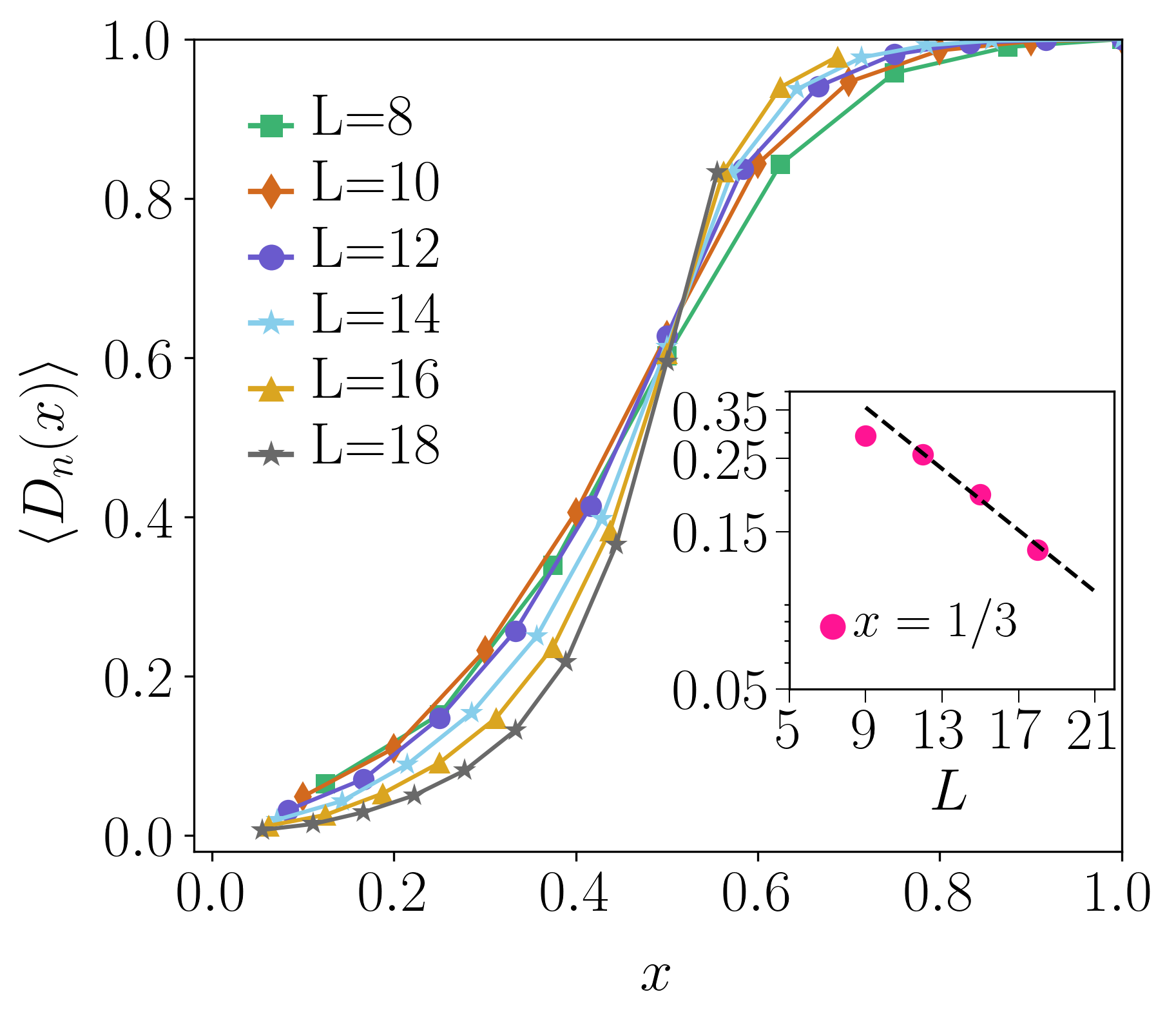}\put(-32,130){$(a)$}\put(-65,110){Chaotic Ising}
\includegraphics[width=53mm]{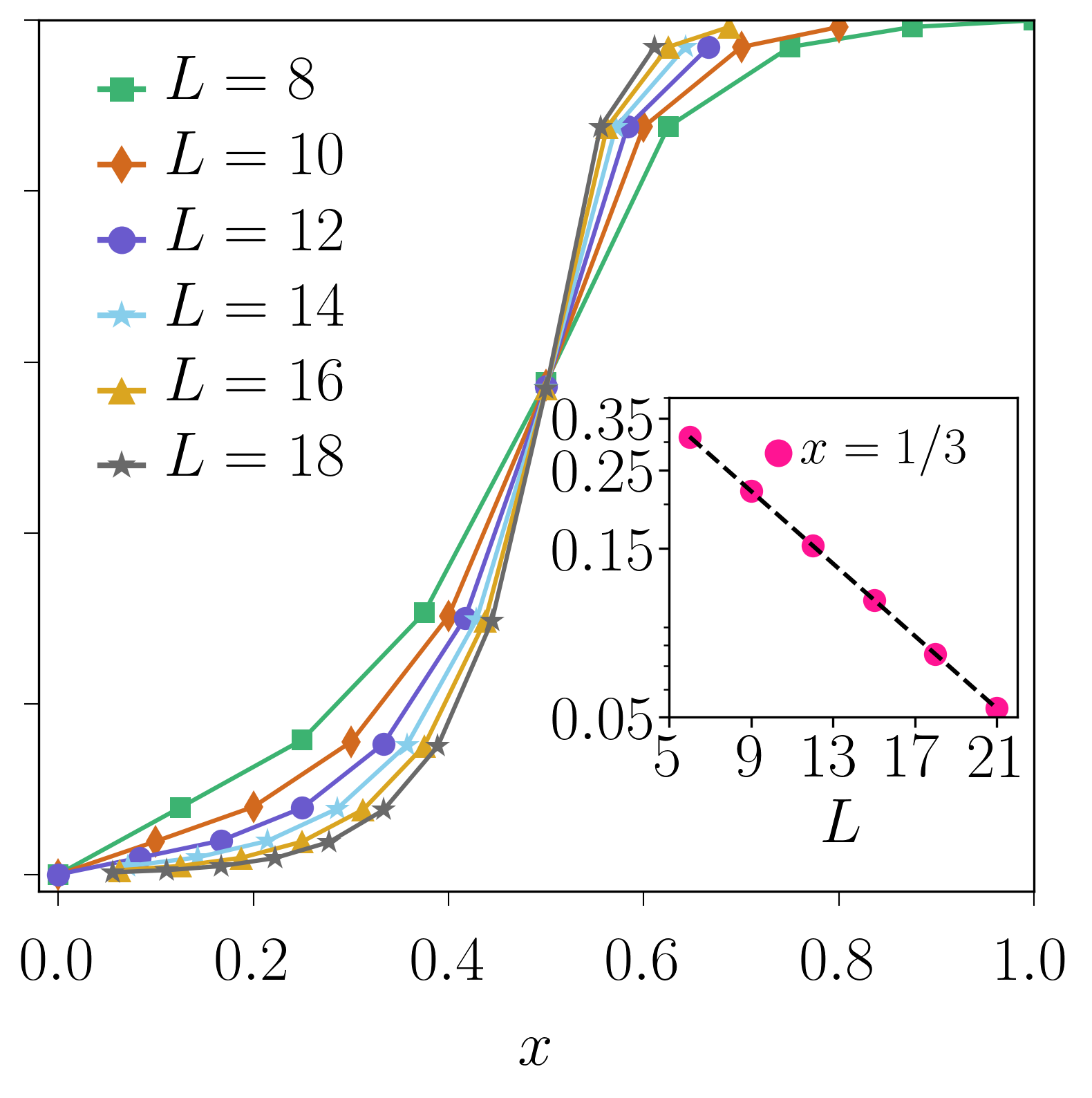}\put(-32,130){$(b)$}\put(-38,113){GOE}	\includegraphics[width=63mm]{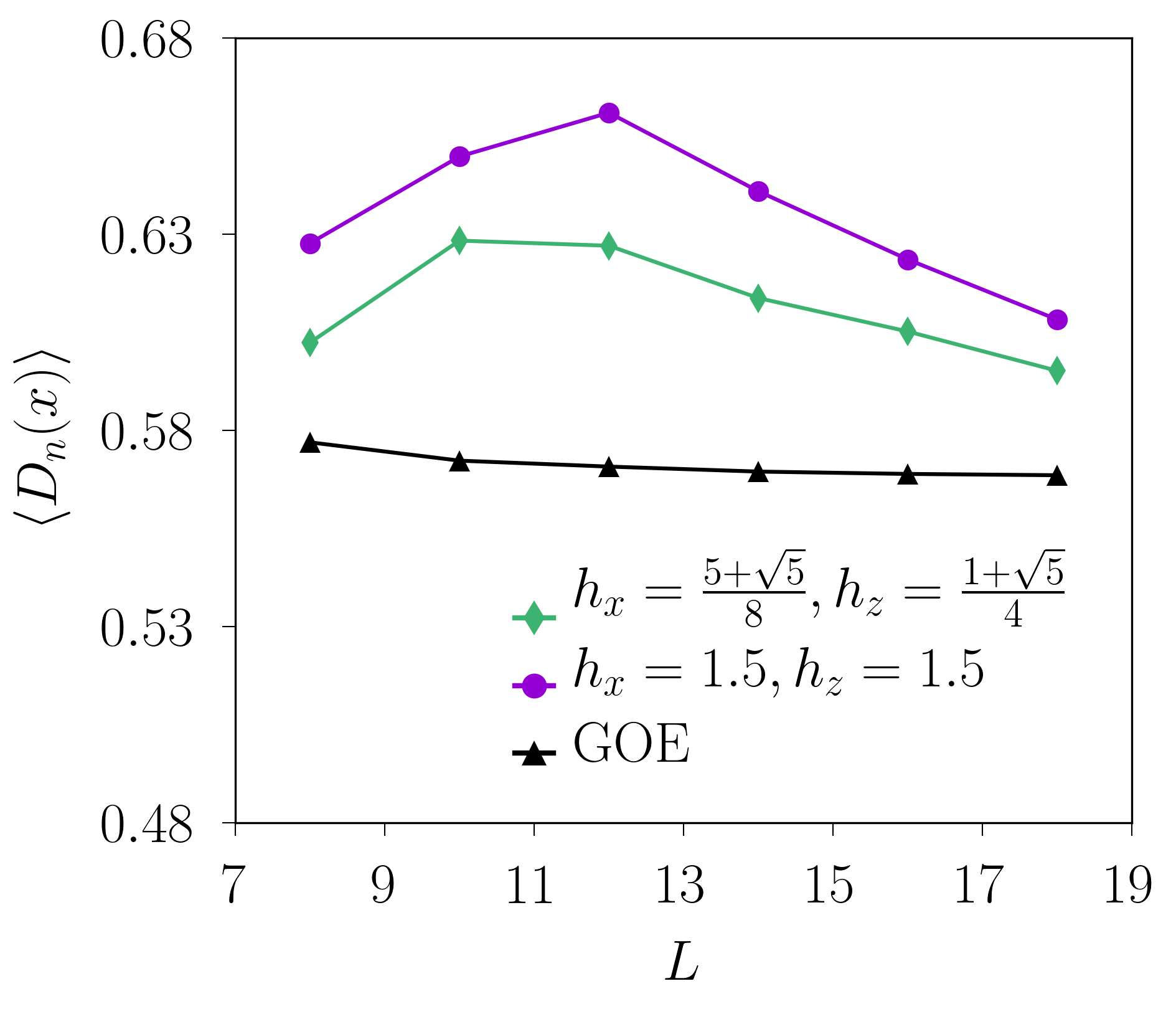}\put(-135,130){$(c)$}\put(-55,130){$x=1/2$}
	\caption{{\bf Quantum many-body chaotic models} : $(a)$ Average $\langle D_{n}(x)\rangle$ for the Ising chain at $h_z=(1+\sqrt{5})/4$ and $h_x=(5+\sqrt{5})/8$ in the symmetry-sector at momentum $K=2\pi/L$. $(b)$ $\langle D_{n}(x)\rangle$ for RMT from the Gaussian orthogonal ensemble (GOE) taken over $4000$ random eigenstates. The inset shows the exponential decay of $\langle D_{n}(x)\rangle$ with system size $L$ for the particular case $x=1/3$. In panel (c), we compare the system-size dependence of $\langle D_{n}(x=1/2)\rangle$ in the quantum many-body chaotic Ising model with RMT.}
	\label{fig:Chaotic_dis_nn}
\end{figure*}

The Ising chain, which is described in Eq.~(\ref{eq:Hamiltonian_Ising}), is known to exhibit chaotic behavior as long as $J$ and $h_{x,z}$ are non-zero~\cite{Huse:2013}. In Fig.~\ref{fig:dis_ave_comp}, we show the qualitatively different behaviors of the distribution of $D_n(x)$ in the integrable and quantum chaotic regimes for a system of size $L=18$ and for $x=1/2,2/9$, respectively. The distribution is narrow and strongly peaked for the quantum chaotic case, and we find that the width of the distribution is exponentially suppressed with the system size in this limit, see Fig.~$S1(a)$ in the Supplemental Material \cite{Supplement}. This result is consistent with RMT, where the distribution is Gaussian, with a vanishing standard deviation by increasing the size of the system, see Fig.~$S1(a,b)$ in the Supplemental Material \cite{Supplement}. Turning off either the longitudinal or transverse field gives an integrable model. We see that these limits have a drastically broader distribution, as shown by the brown and purple curves in Fig.~\ref{fig:dis_ave_comp}.

In what follows, we will consider as a detector of quantum many-body chaos and integrability the mean value $\langle D_{n}(x)\rangle$ taken over all eigenstates in a single momentum sector. In the chaotic regime, we see that $\langle D_{n}(x)\rangle$ decays upon increasing system size $L$ as long as $x<1/2$, see Figs.~\ref{fig:Chaotic_dis_nn}a,b for the chaotic Ising model and the Gaussian Orthogonal Ensemble (GOE) from Random Matrix Theory (RMT). As we show analytically in Sec. \RNum{2} of the Supplemental Material \cite{Supplement}, ETH predicts the following behavior:
\begin{equation}\label{eq:analytic_dis}
\langle D_{n}(x)\rangle=C\Delta~,
\end{equation}
where $\Delta=\langle\epsilon_{n+1}-\epsilon_n\rangle$ denotes the mean level spacing and $C$ a prefactor, which will be discussed in more detail below. Since $\Delta$ decays exponentially with system size $L$ for large systems, we find that this is consequently also the case for $D_{n}(x)$. This holds as long as ETH applies to the considered model system, requiring that the system size $L$ is sufficiently large. From the insets of Figs.~\ref{fig:Chaotic_dis_nn}a,b at a fixed $x=1/3$ we further see that the GOE exhibits compelling evidence for an exponential decay, as predicted by Eq.~(\ref{eq:analytic_dis}). For the chaotic Ising model the behavior is less pronounced due to limitations from the accessible system sizes, but also consistent with the predicted exponential decay.

As we further show analytically in Sec. \RNum{2} of the Supplemental Material \cite{Supplement}, we can additionally bound the proportionality constant $C$. For that purpose we have to assume in addition to ETH that the subsystem density matrix itself can be approximated well by a canonical density matrix. This requires to take the usual limits of statistical mechanics requiring the subsystem to be very large, but still much smaller than the total system. In this regime we find that $C\leq1/\Delta E$ with $\Delta E = \sqrt{\langle H^2 \rangle - \langle H \rangle^2}$ the total energy fluctuations yielding a system-size dependence proportional to $1/\sqrt{L}$. Notice, however, that for the system sizes considered in Fig.~\ref{fig:Chaotic_dis_nn} the aforementioned requirements are not yet met. Thus, the derived bound on $C$ should be viewed more as an asymptotic behavior. 

In Fig.~\ref{fig:Chaotic_dis_nn}, we show numerical data of $\langle D_{n}(x)\rangle$ for many-body quantum chaotic models. 
We find that $x=1/2$ is a fixed point, and the behavior of $\langle D_{n}(x)\rangle$ is different for values of $x$ less than and greater than $1/2$. For $x<1/2$, $\langle D_{n}(x)\rangle$ decays upon increasing $L$, ultimately tending towards zero as $L$ approaches infinity. These results align with the recent study~\cite{Miranda:2023}, which carried out analytical assessments of $\langle D_{n}(x)\rangle$ for random states and has delved into its application in the Ising model for small sizes. 
For $x>1/2$, $\langle D_{n}(x)\rangle$ tends towards 1 by increasing system size. To emphasize the consistency of the results of the chaotic Ising chain with RMT, we study in more detail $\langle D_{n}(1/2)\rangle$ in Fig. \ref{fig:Chaotic_dis_nn}(c). One can see that for large system sizes $\langle D_{n}(1/2)\rangle$ in the chaotic Ising model approaches the saturation value of RMT. More details about RMT are
discussed in Sec. \RNum{1} of the Supplemental Material \cite{Supplement}.

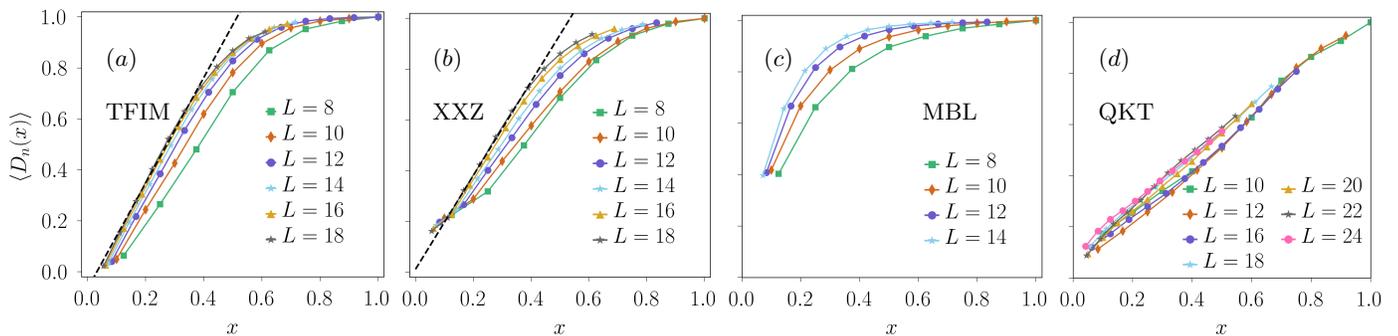
\begin{figure*}[htp]
    \includegraphics[width=52mm]{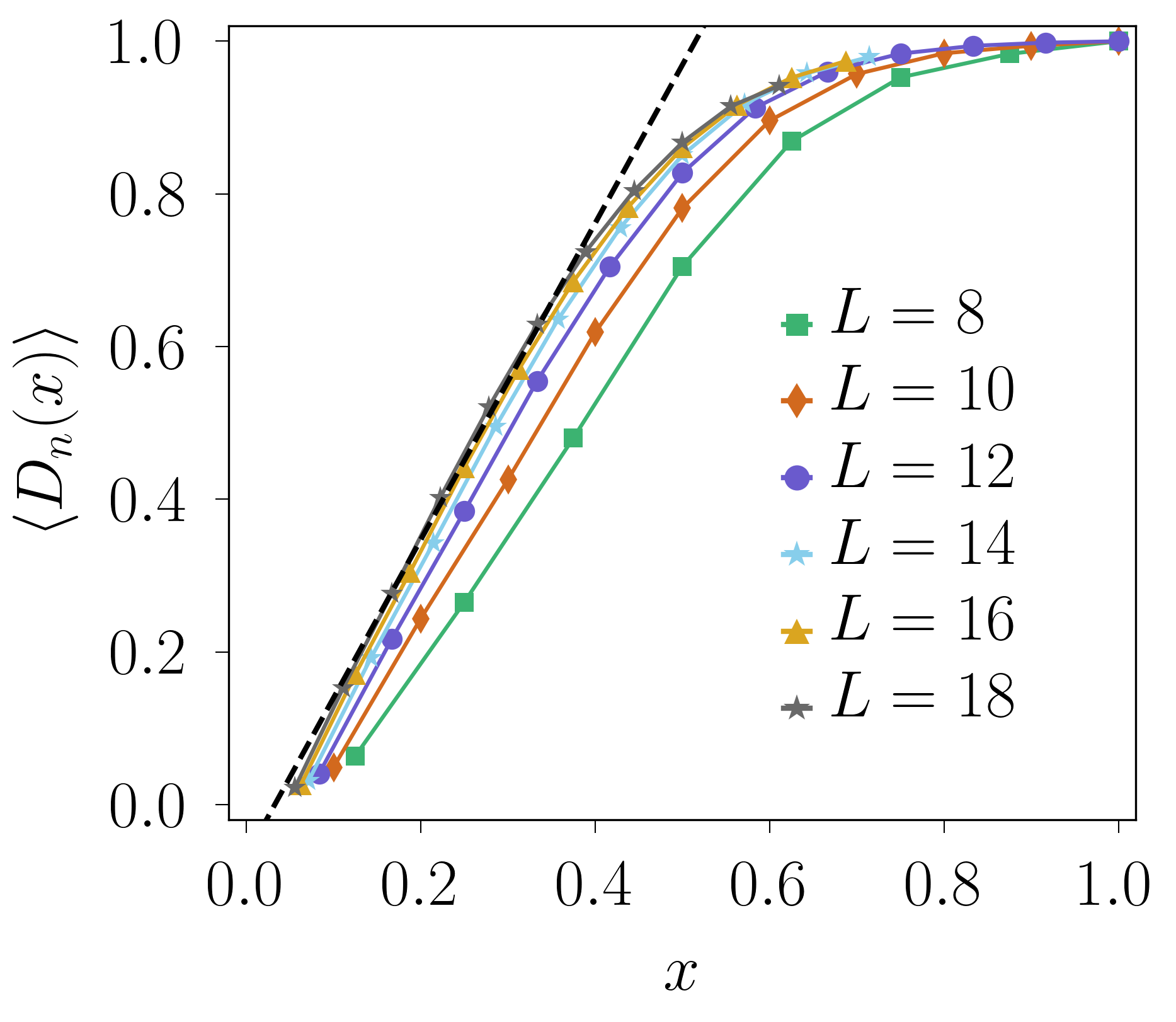}\put(-110,105){$(a)$}\put(-110,85){TFIM}
	\includegraphics[width=43.5mm]{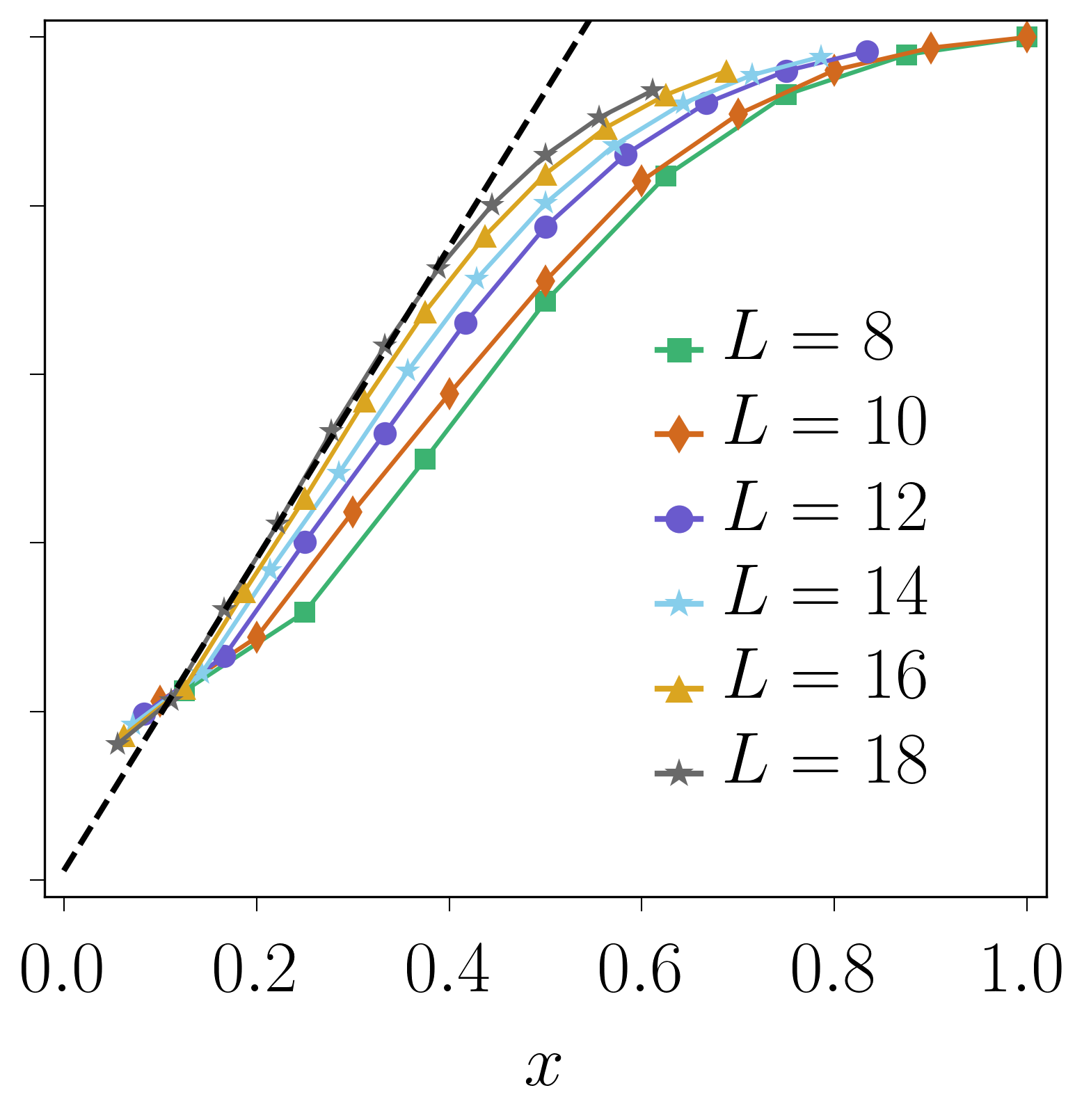}\put(-110,105){$(b)$}\put(-110,85){XXZ}
	\includegraphics[width=44mm]{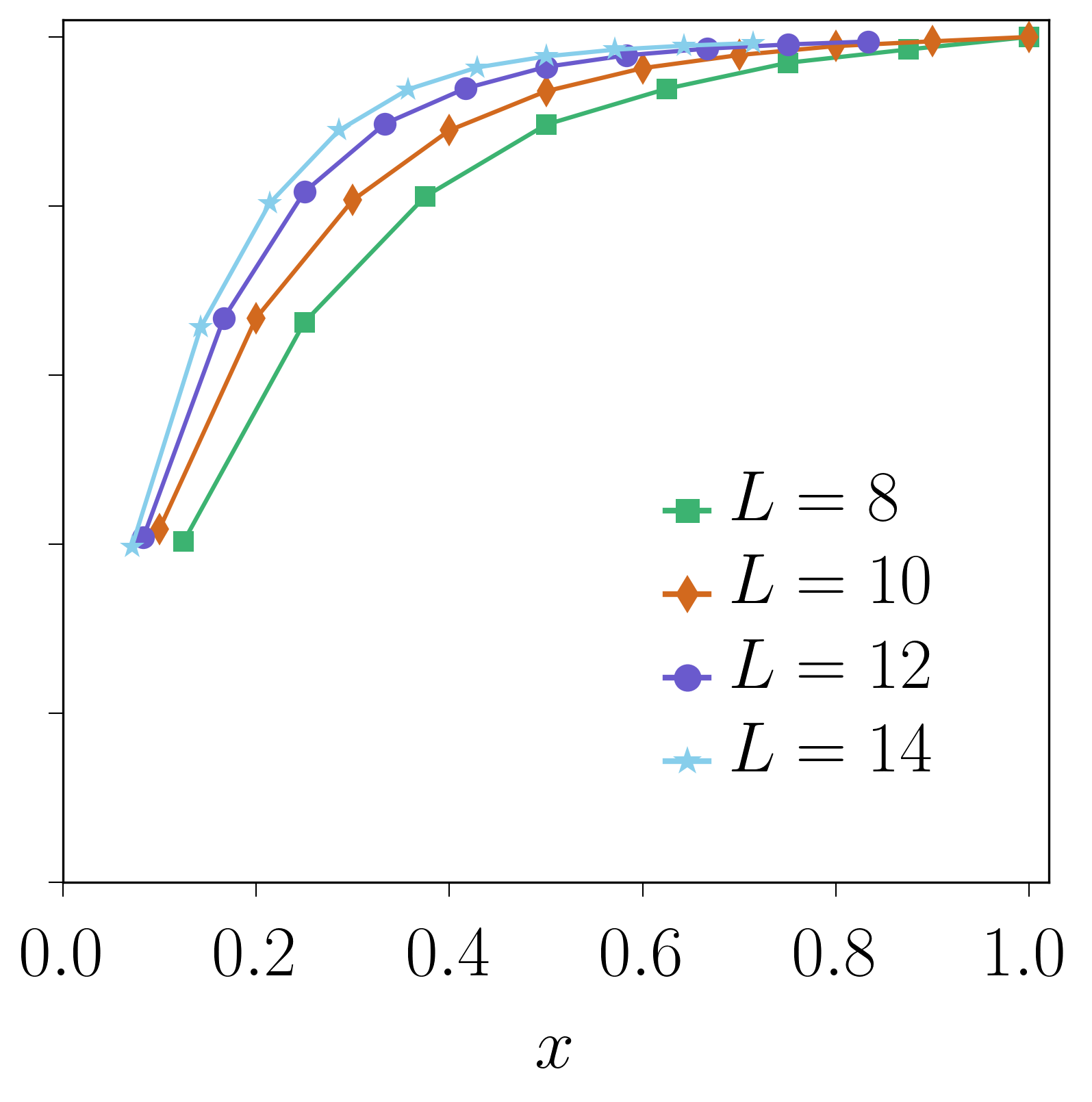}\put(-110,105){$(c)$}\put(-50,85){MBL}
	\includegraphics[width=44.5mm]{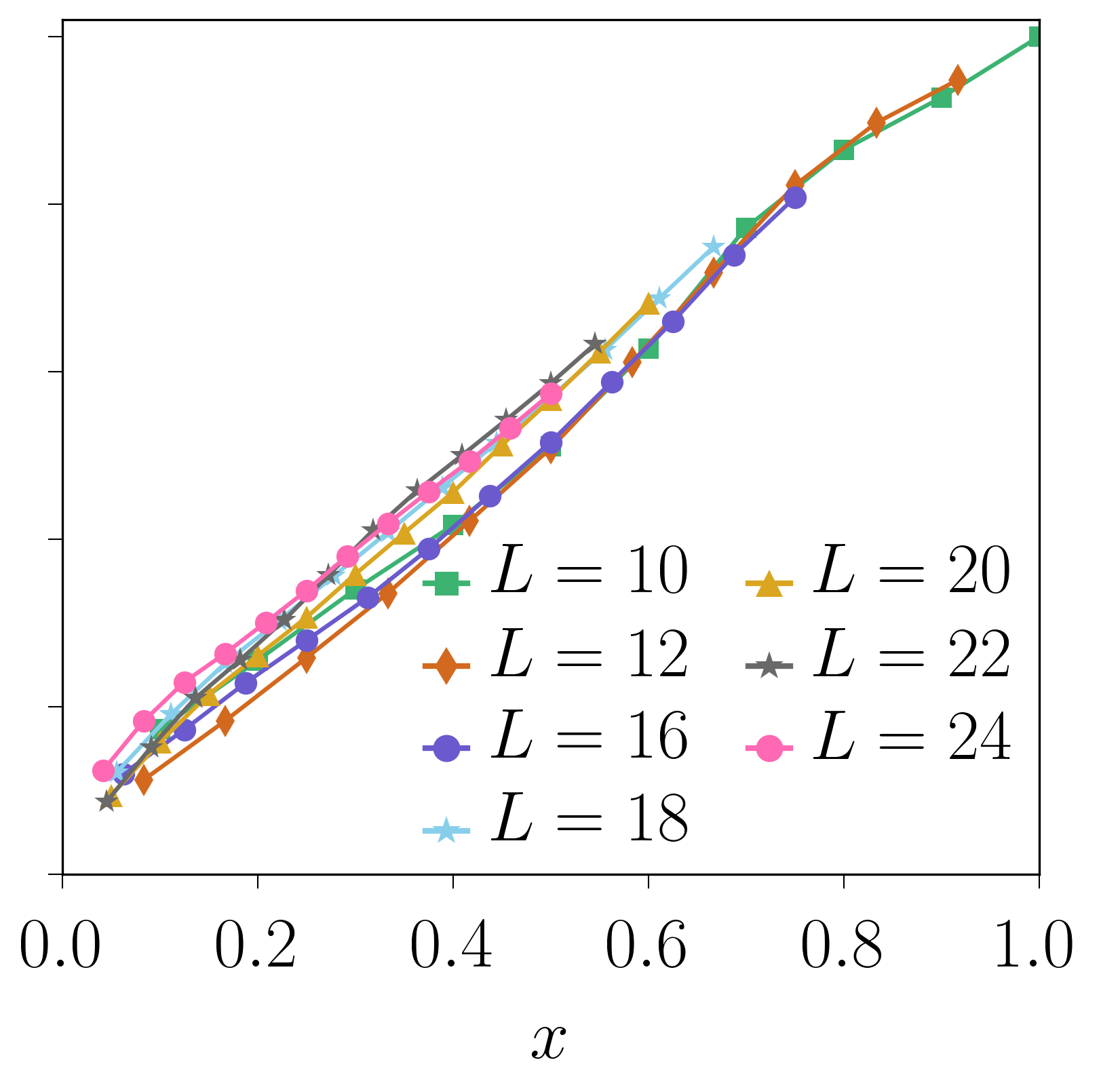}\put(-110,105){$(d)$}\put(-110,85){QKT}
	\caption{{\bf Quantum many-body integrable models}: (a) $\langle D_{n}(x) \rangle$ as a function of $x=L_A/L$ for the transverse field Ising model (TFIM) with $h_x=(5+\sqrt{5})/8$ and $h_z=0$. The dashed line marks the linear $ax$ with $a\approx 2$. (b) $\langle D_{n}(x) \rangle$ for the Bethe-ansatz integrable XXZ chain with $\Delta=2$. Again, we have included a dashed line indicating linear behavior $ax$ with $a\approx 2$.  (c) XXZ model with $\Delta =2$ and strong disorder $h=10$ in the MBL regime. (d) Quantum kicked top (QKT) model at $\kappa =7$ and $\tau=1$. Here in (a,b,c) we focus on the symmetry block with $K = 2\pi/L$, and in (d) the average is taken over the $S=L/2$ sector.}
	\label{fig:Integrable_models}
\end{figure*}

{\it{Integrable systems}}: After discussing the behavior of $\langle D_{n}(x)\rangle$ for quantum many-body chaotic systems, we now move to the case of integrable models. In particular, we first study the transverse-field Ising chain at $h_z=0$ in Eq.~(\ref{eq:Hamiltonian_Ising}). Since the model is integrable in this regime, we expect the results to be significantly different from those predicted by RMT. We show the markedly different behavior in Fig.~\ref{fig:Integrable_models}(a) where we provide numerical data for $\langle D_{n}(x)\rangle$ for different system sizes. 
First of all, we can not identify a fixed point anymore. Upon increasing system size, we find that $\langle D_{n}(x)\rangle$ appears to converge to a single non-zero curve that does not match the prediction of ETH in Eq.~\ref{eq:analytic_dis}. Instead, the numerical data seems to approach a linear function $\langle D_{n}(x)\rangle\sim ax$ upon increasing system size with a slope $a\approx 2$. The linear slope remains roughly unchanged when considering other symmetry blocks except those with $K=0,\pi$, which have $a\approx 1$ \cite{footnote1}. This is discussed further in Sec. \RNum{3} of the Supplemental Material \cite{Supplement}. 

In order to gain some analytical insights into this linear behavior, we have also studied the XY chain, which can be solved exactly through a mapping to a free fermion Hamiltonian using the Jordan-Wigner transformation~\cite{Lieb:1961fr,Katsura:1962hqz,Pfeuty:1970ayt}. The subsystem trace distance between two eigenstates can be obtained by explicit construction of the RDMs~\cite{Zhang:2019wqo,Zhang:2019itb,Zhang:2022tgu}. We find evidence that $\lag D_{n}(x) \rag \propto x$ in the large $L$ limit, especially in the range $x\in(0.1,0.4)$, see Sec. \RNum{4} of the supplemental material \cite{Supplement}.

In the following, we will now explore to what extent this observed behavior for the Ising chain generalizes also to other models. For that purpose, we will conduct the same analysis on several different paradigmatic examples of integrable systems, including the Bethe-ansatz integrable homogeneous XXZ chain, the XXZ model with a random field in the MBL regime, and the quantum many-body kicked top. 

The Hamiltonian of the XXZ model is given by 
\begin{equation}\label{eq:Hamiltonian_XXZ}
H_{XXZ}=\sum_{l=1}^{L}(\sigma_l^x\sigma_{l+1}^x+\sigma_l^y\sigma_{l+1}^y+\Delta\sigma_l^z\sigma_{l+1}^z+h_z^l\sigma_l^z)~,
\end{equation}
where $\Delta$ denotes the anisotropy parameter. We will consider two cases for the magnetic field $ h_z^l$. On the one hand, we take $h_z^l=h_z$ uniform, where Eq. (\ref{eq:Hamiltonian_XXZ}) yields the Bethe-ansatz integrable XXZ model. On the other hand, we will consider $h_z^l \in [-h,h]$ a random variable drawn from the uniform distribution, which gives a paradigmatic model for MBL~\cite{Marco:2008,Huse:2014}.

In Fig. \ref{fig:Integrable_models} (b), we show numerical data of $\langle D_{n}(x)\rangle$ for the Bethe-ansatz integrable limit. We observe a similar behavior as compared to the integrable Ising chain. Again, $\langle D_{n}(x)\rangle$ doesn't align with the ETH prediction of Eq.~(\ref{eq:analytic_dis}), but rather approaches a linear behavior as a function of $x$ with a slope $a\approx2$ as in the case of the Ising chain.

For the case of strong random fields, the XXZ model enters an MBL phase~\cite{Marco:2008,Huse:2014}, violating ETH. Systems in the fully MBL regime are expected to show an emergent form of integrability caused by the presence of an extensive number of emergent local conservation laws. The numerical results for the MBL case, see Fig. \ref{fig:Integrable_models} (c), again don't follow the ETH prediction in Eq.~(\ref{eq:analytic_dis}) so that our indicator correctly classifies such MBL systems as integrable. However, differently from the other integrable systems, we don't observe a linear behavior of $\langle D_{n}(x)\rangle$ as a function of $x$. $\langle D_{n}(x)\rangle$ rather appears to approach $\langle D_{n}(x)\rangle \to 1$ upon increasing system size.

Finally, we investigate a many-body Floquet system whose dynamics is captured by the following Hamiltonian~\cite{haake:1987}
\begin{equation}\label{Ising_kicked}
H= h_x\sum_{l=1}^{L} \sigma_l^x+\frac{\kappa}{L}\sum_{l,m=1}^{L}\sigma^z_l\sigma_m^z\sum_{n=-\infty}^{+\infty} \delta(t-n\tau),  
\end{equation}
where the transverse field and kick strength are represented by $h_x$ and $\kappa$, respectively, and the kicks have a period of $\tau$. This is the so-called quantum kicked top (QKT). In this model, the collective spin operator $S^2=S_x^2+S_y^2+S_z^2$, with $S_{\alpha}=1/2\sum_{i}\sigma_i^{\alpha}$ where $\alpha=x,y,z$, is conserved. As a result, the Hilbert space of the system can be divided into subspaces with fixed total spin, in which the dynamics are equivalent to that of a single kicked top with the corresponding angular momentum $S$~\cite{Wang:2004,Milburn:1999,Dogra:2019}.
We analyze this explicitly time-dependent model using Floquet theory by means of the eigenstates of the Floquet operator $U_F=\exp{(-i\kappa S_z^2/L )}\exp{(-ih_x \tau S_x)}$, i.e., the time-evolution operator $U_F$ over one period $\tau$.
As the analog of the eigenenergies we consider the quasienergies of the corresponding Floquet operator $H_F = -i\ln(U_F)$.
In what follows, we perform the statistics for the largest subspace of Hilbert space, which has a size proportional to system size $L$.

The key consequence of the aforementioned conservation law is that the largest subspace in Hilbert space (i.e. $S=L/2$ sector) is linear in system size $L$. Therefore, the exact solution can be obtained with an effort depending only polynomially on system size. As a consequence, we consider this model to be quantum \textit{many-body} integrable. Let us directly emphasize, however, that this statement is not in contradiction to the well-known result that the kicked-top is considered a paradigmatic quantum chaotic model in the \textit{single-particle} sense. In particular, it has been shown that the level spacing statistics of individual blocks in Hilbert space can display GOE behavior whenever the parameter $\kappa$ exceeds a critical value $\kappa_c$~\cite{Alicki:1996,Wang:2004,Ghose:2008}.
Such GOE behavior is a standard indicator of quantum chaos.
%

%
%
%
%

In Fig.~\ref{fig:Integrable_models}(d) we show the results of $\langle D_{n}(x) \rangle$ for the QKT with $\kappa=7$.
One can clearly observe that $\langle D_{n}(x) \rangle$ doesn't follow the ETH prediction in Eq.~(\ref{eq:analytic_dis}) of an exponentially decaying $\langle D_{n}(x) \rangle$ with system size.
We rather find that $\langle D_{n}(x) \rangle$ behaves similar to the conventional integrable models such as the Ising chain or the XXZ model.
Upon increasing system size, $\langle D_{n}(x) \rangle$ appears to approach a linear behavior upon increasing $L$.
As a consequence, our indicator predicts the many-body quantum integrability of the QKT for $\kappa=7$. It is noteworthy that, for small values of the parameter $\kappa$, the behavior of the observable $D_n(x)$ is unexpectedly consistent with that of a MBL system. Additional illustrations of this can be found in Sec. \RNum{5} of the supplemental material \cite{Supplement}, where we present an analysis of $D_n(x)$ for a range of $\kappa$ values.

%
%

\section{Conclusions}\label{sec:Conclusions} 

In this work, we have presented an alternative indicator for quantum many-body integrability and chaos through trace distances of nearest-neighboring eigenstates, unlike traditional level spacing statistics focusing on Hamiltonian eigenvalues. In recent years many indicators based on local observables related to the reduced density matrix of small subsystems have been introduced. However, according to our analysis in Sec. \RNum{6} of the Supplemental Material~\cite{Supplement} examining the corresponding density matrix of larger subsystems (i.e., $x>0$) proves more fruitful in discerning many-body integrable systems from chaotic regimes.

Trace distances also provide bounds on eigenstate-to-eigenstate fluctuations of some non-local quantities, such as the entanglement entropy and R{\'e}nyi entropy~\cite{Fannes:1973,Audenaert:2007,Chehade:2019}, which have been used as alternative indicators for chaos, integrability, and MBL in quantum systems~\cite{Vidmar:2017,Hackl:2017,Blond:2019}. Let us point out, however, that by utilizing the entanglement entropy, say, as an indicator it is essential to have analytical access to a reference, which is typically the Page value. While this allows to study many-body quantum chaos at infinite temperature for bounded Hamiltonians, the indicator based on trace distances in our work exhibits a broader range of applicability as it can be also applied in an energy-resolved fashion, in principle, and also for unbounded Hamiltonians including bosonic systems.

For the future, it would be important to develop a deeper analytical understanding of our indicator $\langle D_{n}(x) \rangle$ in the quantum many-body integrable regimes, which so far has been mostly based on numerical results.
A particularly interesting point might be to address the seemingly universal linear behavior of $\langle D_{n}(x) \rangle$ for certain integrable models and why certain other models, such as MBL systems apparently behave differently. 
While we have already been covering a broad range of physical systems of different types, it would be certainly important to further explore the applicability of our indicator to an even larger class of models.  
For instance, it would be interesting to apply the indicator to interacting systems with disorder, where recent developments have raised fundamental questions on the MBL phase and the MBL transition~\cite{Suntajs:2020,Abanin:2021,Sels:2021,Roeck:2017}.
In this context, it might be a natural question to which extent our indicator might be well suited to extract the MBL transition and its system-size dependence.

\section*{Acknowledgements}\label{sec:Acknowledgements}
We acknowledge valuable discussions with A. Russomanno. This project has received funding from the European Research Council (ERC) under the European Union’s Horizon 2020 research and innovation program (Grant Agreement No. 853443). MAR thanks CNPq for partial support. JZ thanks support from the National Natural Science Foundation of China (NSFC) grant number 12205217.%

\hfill\\
\newpage
\providecommand{\href}[2]{#2}\begingroup\raggedright

\endgroup
\clearpage

\widetext
\begin{center}
\textbf{\large Supplemental Materials: Identifying quantum integrability and many-body quantum chaos using eigenstates
trace distances}
\end{center}
\setcounter{equation}{0}
\setcounter{figure}{0}
\setcounter{table}{0}
\setcounter{section}{0}
\setcounter{page}{1}
\makeatletter
\renewcommand{\theequation}{S\arabic{equation}}
\renewcommand{\thefigure}{S\arabic{figure}}
\renewcommand{\bibnumfmt}[1]{[S#1]}
\renewcommand{\citenumfont}[1]{S#1}

\section{Random matrix Theory}\label{app:RMT}
The statistical characteristics of the spectrum of a chaotic system are often described by random matrices with Gaussian distributions~\cite{Guhr:1998,Forrester:2003}. In this supplementary material, we focus on two specific symmetry classes of these random ensembles: GOE (Gaussian Orthogonal Ensemble) and GUE (Gaussian Unitary Ensemble). The spectral statistics of these two ensembles remain unchanged under the transformations of orthogonal and unitary matrices, respectively. Both contain square matrices with real eigenvalues, but the entries of the matrices can either be real (GOE) or complex (GUE) independent random variables. Level statistics of random matrices in Gaussian ensembles are studied in different references; for a review, see~\cite{Livan:2018}. The distribution of the consecutive level spacing (i.e. $s_n=\epsilon_{n+1}-\epsilon_{n}$) in these Gaussian ensembles follows the Winger-Dyson statistics, $p(s)\sim s^{\beta}e^{-C(\beta)s^2}$, where $\beta=0$ and $\beta=1$ correspond to GOE and GUE, respectively. In~\cite{Atas:2018}, the authors also consider the distribution of the ratios of level spacing $r_n=s_n/s_{n-1}$, which has the same level repulsion as $p(s)$ for small $r$ (i.e., $p(r)\sim r^{\beta}$) and follows $p(r)\sim r^{-(2+\beta)}$ for large $r$. It is important to note that the statistics of spin models follow these random ensembles depending on the presence or absence of time-reversal symmetry (TRS). GOE is used to describe systems with TRS, while GUE can be applied in the absence of this symmetry.

\begin{figure*}
	\includegraphics[width=62mm]{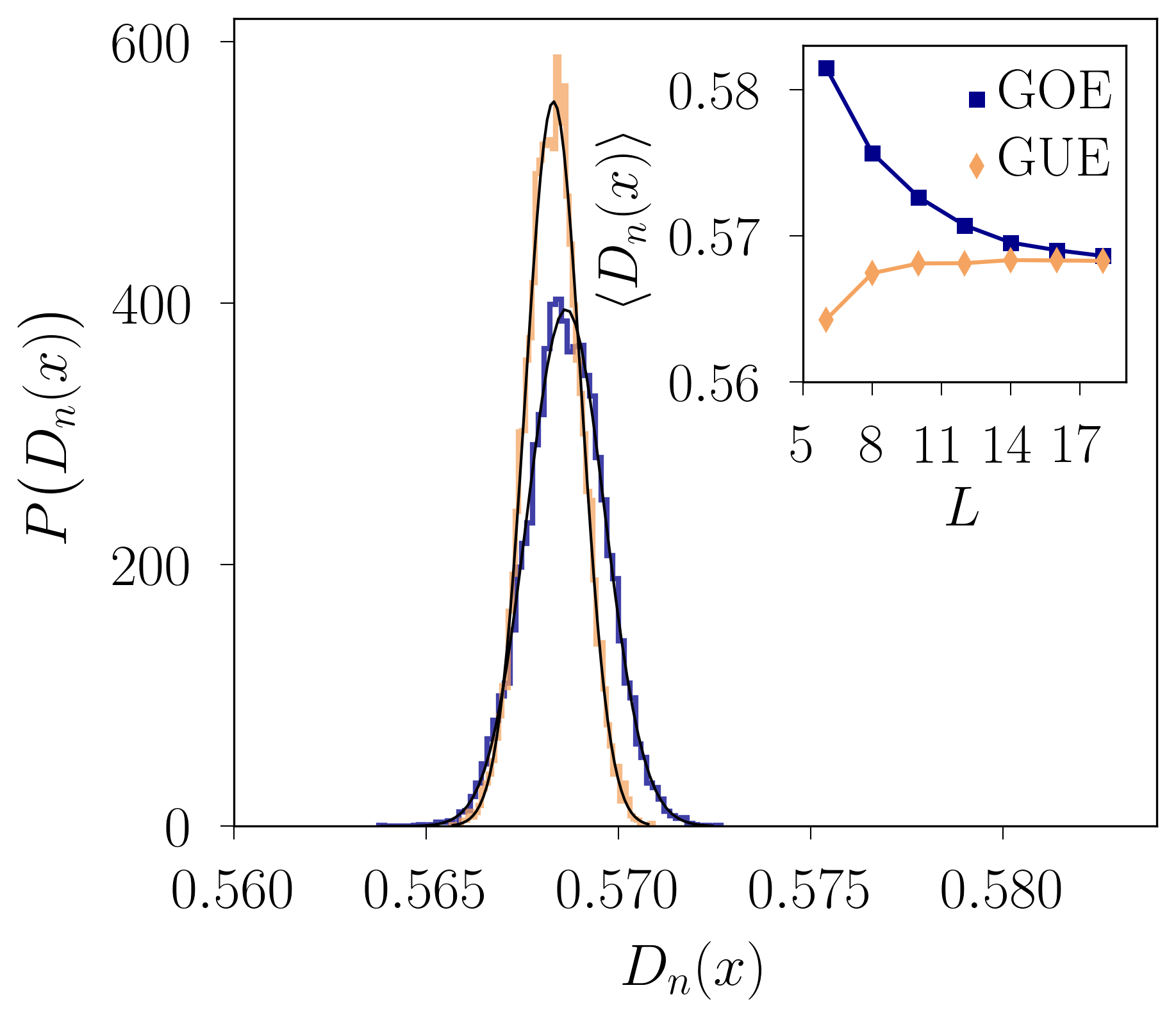}\put(-138,140){$(a)$}\put(-134,100){$x=1/2$}\hspace{4mm}
	\includegraphics[width=65mm]{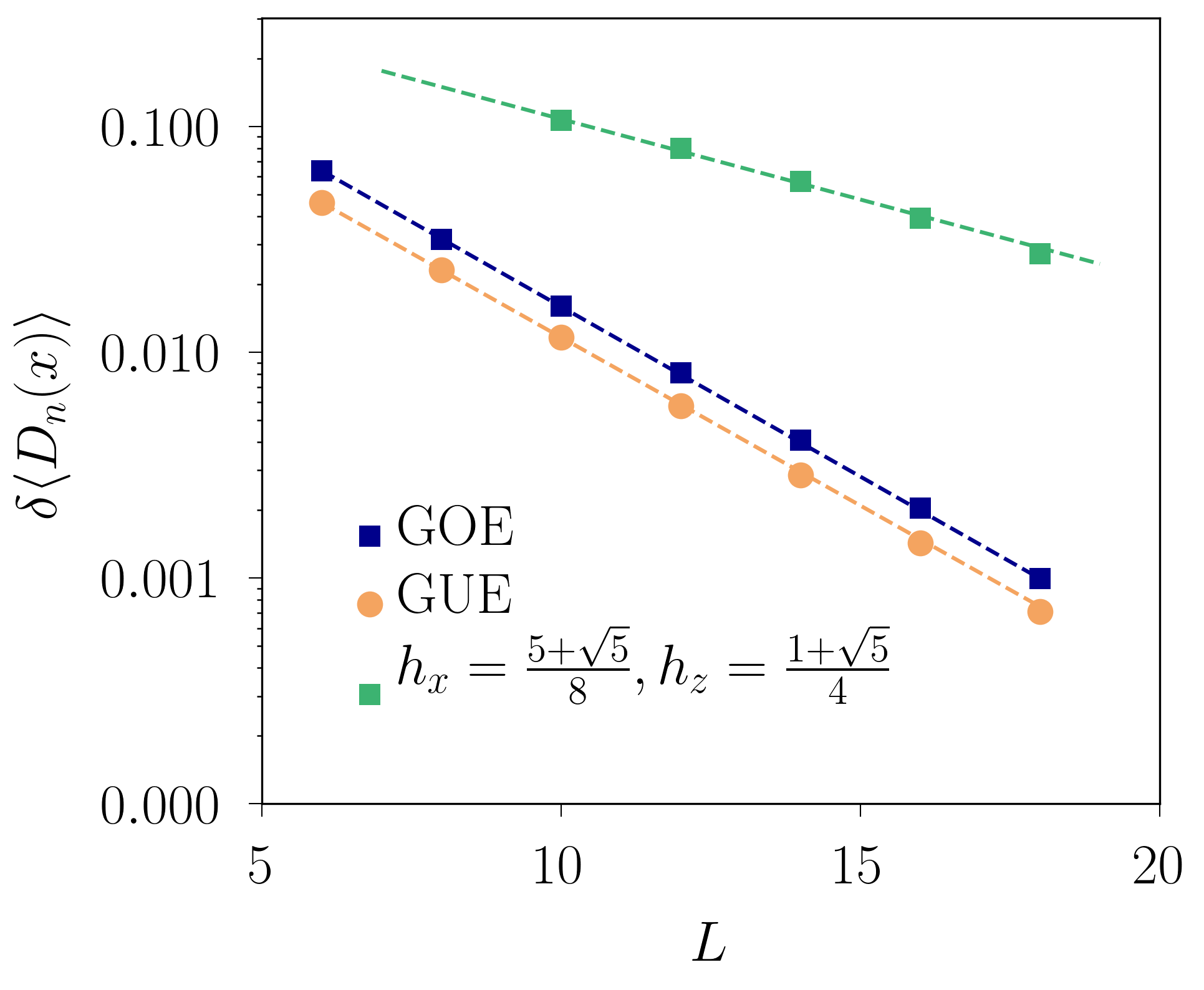}\put(-30,140){$(b)$}\put(-70,100){$x=1/2$}
	\caption{{\bf RMT}: (a) Gaussian distribution of eigenstates trace distances for $L=18$. The blue curve represents GOE with fitting parameters $\mu=0.568,\sigma=0.001$, and the orange one belongs to GUE with $\mu=0.568,\sigma=0.0007$. The inset in the top right shows the convergence of $\langle D_{n}(0.5)\rangle$ for GUE and GOE to the same value. $(b)$ The standard deviation of distance distribution decreases exponentially $(\sim e^{-bL})$ with system size. The fitting exponents are
	$b\sim 0.34$ for both GOE and GUE and $b\sim 0.17$ for chaotic Ising model.}
	\label{fig:RMT_dis}
\end{figure*}

The focus of this study is to explore the significance of random matrix theory (RMT) in chaotic spin models from a novel perspective. We investigate the statistical properties of the trace distances between the eigenstates of chaotic Hamiltonians and compare them to those of random matrix ensembles. This can be achieved through the use of either random matrix ensembles~\cite{Guhr:1998} or the construction of an ensemble of random states~\cite{Miszczak:2012}. The numerical results are presented in Fig.~\ref{fig:RMT_dis}. In Fig.~\ref{fig:RMT_dis}(a), we show the distribution of $D_{n}(x)$ where $x=1/2$. The distribution of eigenstate spacings seems to follow a Gaussian law for both GUE and GOE. The standard deviation of the distributions decays exponentially with the size of the system ($\sim e^{-\eta L}$); see Fig.~\ref{fig:RMT_dis}(b). Therefore, the Gaussian distribution converges to a delta function peak in the thermodynamic limit. We see similar behavior for the chaotic Ising model, as shown by the green points in Fig.~\ref{fig:RMT_dis}(b), but the decay is slower. As we can see from the Fig.~\ref{fig:RMT_dis}(a) subplot, the average of $D_{n}(x)$ converges to the same value for both Gaussian ensembles as the system size increases.
\section{Eigenstates trace distances in the canonical ensemble}\label{app:analytic_ETH}
In this section, we demonstrate that the exponential decay of $D_{n}(0<x<0.5)$ with increasing system size observed in a general chaotic Hamiltonian, $\big($see Fig.~2(b)$ \big)$, is consistent with the eigenstate thermalization hypothesis (ETH) using analytical calculation in the canonical ensemble. 

\subsection{Derivation of Eq.5}

For any observable that satisfies the ETH, the expectation value of the observable with respect to a finite energy density eigenstate $\ket{\psi_n}$ is identical to that derived from a thermal ensemble. If this is true for all operators defined in a subsystem $A$, then the reduced density matrix $\rho_n^A=\Tr_{\bar{A}}\ket{\psi_n}\bra{\psi_n}$ is equivalent to the corresponding thermal reduced density matrix
\begin{equation}\label{eq:reduced_n}
\rho_{n,th}^A=\Tr_{\bar{A}}\frac{e^{-\beta(\epsilon_n)H}}{Z\big(\beta(\epsilon_n)\big)}~,    
\end{equation}
where $\beta^{-1}$ is temperature, one can assign it to each Hamiltonian eigenstate such that the energy expectation value in the canonical ensemble is equal to the eigenstate energy. It has been observed that when the size of the subsystem $L_A$ is fixed and $L\to \infty$, ETH holds for all observables defined in $A$, and thus eigenstate and thermal reduced density matrices become equal. However, in the general case where $x<0.5$, this is not true for all observables as ETH fails for some, such as those based on conserved quantities of the system. The number of such operators is exponentially smaller than the total number of independent operators in a subsystem. In this case, $\rho_n^A$ and $\rho_{n,th}^A$ are not equal, but they still provide the same expectation value for a large number of observables~\cite{S:Grover:2018}. In the following, we aim to demonstrate that the trace distances of the canonical reduced density matrices, when restricted to subsystem $A$, behave similarly to those of the eigenstate matrices. One can rewrite Eq.~\ref{eq:reduced_n} in terms of free energy, $F\big(\beta(\epsilon_n)\big)=-\ln{ Z\big(\beta(\epsilon_n)\big)}/\beta(\epsilon_n)$, so we have
\begin{equation}\label{eq:reduced_n_F}
\rho_{n,th}^A=\Tr_{\bar{A}} e^{-\beta(\epsilon_n)\Big(H-F\big(\beta(\epsilon_n)\big)\Big)}~.    
\end{equation}
According to ETH, the nearby eigenstates have a small energy difference. Therefore one can expand $\beta(\epsilon_{i+1})$ in terms of energy differences as
\begin{equation}\label{eq:temp_expan}
\beta(\epsilon_{n+1})=\beta(\epsilon_{n})+\frac{\Delta\epsilon_n}{K_BT^2\eta}+{\cal O}(\Delta\epsilon_n^2)~.    
\end{equation}
Here $\Delta\epsilon_n=\epsilon_{n+1}-\epsilon_{n}$, and $\eta=\partial E/\partial T|_{E=\epsilon_{n}}$ is heat capacity. Considering Eq.~\ref{eq:temp_expan}, the free energy at inverse temperature $\beta(\epsilon_{n+1})$ is expanded as  
\begin{equation}\label{eq:free_expan}
\begin{split}
F\big(\beta(\epsilon_{n+1})\big)= & F\big(\beta(\epsilon_{n})+\frac{\Delta\epsilon_n}{K_BT^2\eta}+{\cal O}(\Delta\epsilon_n^2)\big)= F\big(\beta(\epsilon_n)\big)+\\&\frac{\partial F\big(\beta(\epsilon_n)\big)}{\partial\beta(\epsilon_n)}\frac{\Delta\epsilon_n}{K_BT^2\eta}+{\cal O}(\Delta\epsilon_n^2)~.
\end{split}
\end{equation}
By replacing Eq.~\ref{eq:temp_expan} and Eq.~\ref{eq:free_expan} in Eq.~\ref{eq:reduced_n_F} and after some simplification the reduced density matrix corresponding to ($n+1$)-th eigenstate can be estimated as
%

\begin{equation}\label{eq:rho_i+1}
\rho_{n+1,th}^A=\Tr_{\bar{A}}e^{-\beta(\epsilon_{n})\Big(H-F\big(\beta(\epsilon_{n})\big)\Big)}\Bigg[1-\Big(\frac{\partial F\big(\beta(\epsilon_{n})\big)}{\partial\beta(\epsilon_n)}\beta-H+F\big(\beta(\epsilon_{n})\big)\Big)\frac{\Delta\epsilon_n}{K_BT^2\eta}\Bigg]+{\cal O}(\Delta\epsilon_n^2)=\rho_{n,th}-\frac{\partial\rho_n}{\partial\beta(\epsilon_{n})}\frac{\Delta\epsilon_n}{K_BT^2C}+{\cal O}(\Delta\epsilon_n^2). 
\end{equation}
%

By moving the $\rho_n$ to the left-hand side and taking the norm from both sides, we have
\begin{equation}\label{eq:distance}
||\rho_{n+1,th}^A-\rho_{n,th}^A||= C\Delta\epsilon_n+{\cal O}(\Delta\epsilon_n^2) ~.  
\end{equation}
where $C(\epsilon_{n})=||\frac{\partial\rho_n^A}{\partial\beta(\epsilon_{n})}||\frac{1}{K_BT^2\eta}.$ Eq.~5 in the main text is derived from averaging distances of all nearest-neighbor eigenstates. Under the ETH assumption, $C(\epsilon_{n})$ presents a smooth function of $\epsilon_{n}$, leading to averaging solely over $\Delta\epsilon_n$.
\subsection{Define an upper bound for $C$}

In the thermodynamic limit and fixed ratio $x$ we have $\rho_{n,th}^A=e^{-\beta(\epsilon_n) H_A}/Z_A$~\cite{S:Dymarsky:2016ntg}. The derivative of $\rho_n^A$ in respect to $\beta(\epsilon_n)$ has the following form
\begin{equation}\label{eq:der_rhoA}
||\frac{\partial\rho_n^A}{\partial\beta(\epsilon_n)}||=||(E_A-H_A)\frac{e^{-\beta(\epsilon_n) H_A}}{Z_A}||~.  
\end{equation}
By considering $H_A\ket{E_{\lambda}}=E_{\lambda}\ket{E_{\lambda}}$, we rewrite the right hand side of Eq.~\ref{eq:der_rhoA} in the base of eigenstates of $H_A$
\begin{equation}\label{eq:der_rhoA_lam}
||\frac{\partial\rho_n^A}{\partial\beta(\epsilon_n)}||=||\sum_{\lambda}p_{\lambda}(E_A-E_{\lambda})\ket{E_{\lambda}}\bra{E_{\lambda}}||~,  
\end{equation}
where $p_{\lambda}=\frac{e^{-\beta(\epsilon_n) E_{\lambda}}}{Z_A}$ is the probability of being in configuration $\ket{\lambda}$. The right hand side of Eq.~\ref{eq:der_rhoA_lam} has the following upper bound : $\sum_{\lambda}|E_{A}-E_{\lambda}|p_{\lambda}\leq \sqrt{\sum_{\lambda}(E_A-E_{\lambda})^2 p_{\lambda}}=\sqrt{\braket{H_A^2}-\braket{H_A}^2}=\Delta E$, where $\Delta E$ is energy fluctuations. Implementing that in Eq.\ref{eq:der_rhoA_lam} gives
\begin{equation}\label{eq:def_rho}
||\frac{\partial\rho_n^A}{\partial\beta(\epsilon_n)}||\leq\Delta E.    
\end{equation}
\begin{figure*}
	\includegraphics[width=65mm]{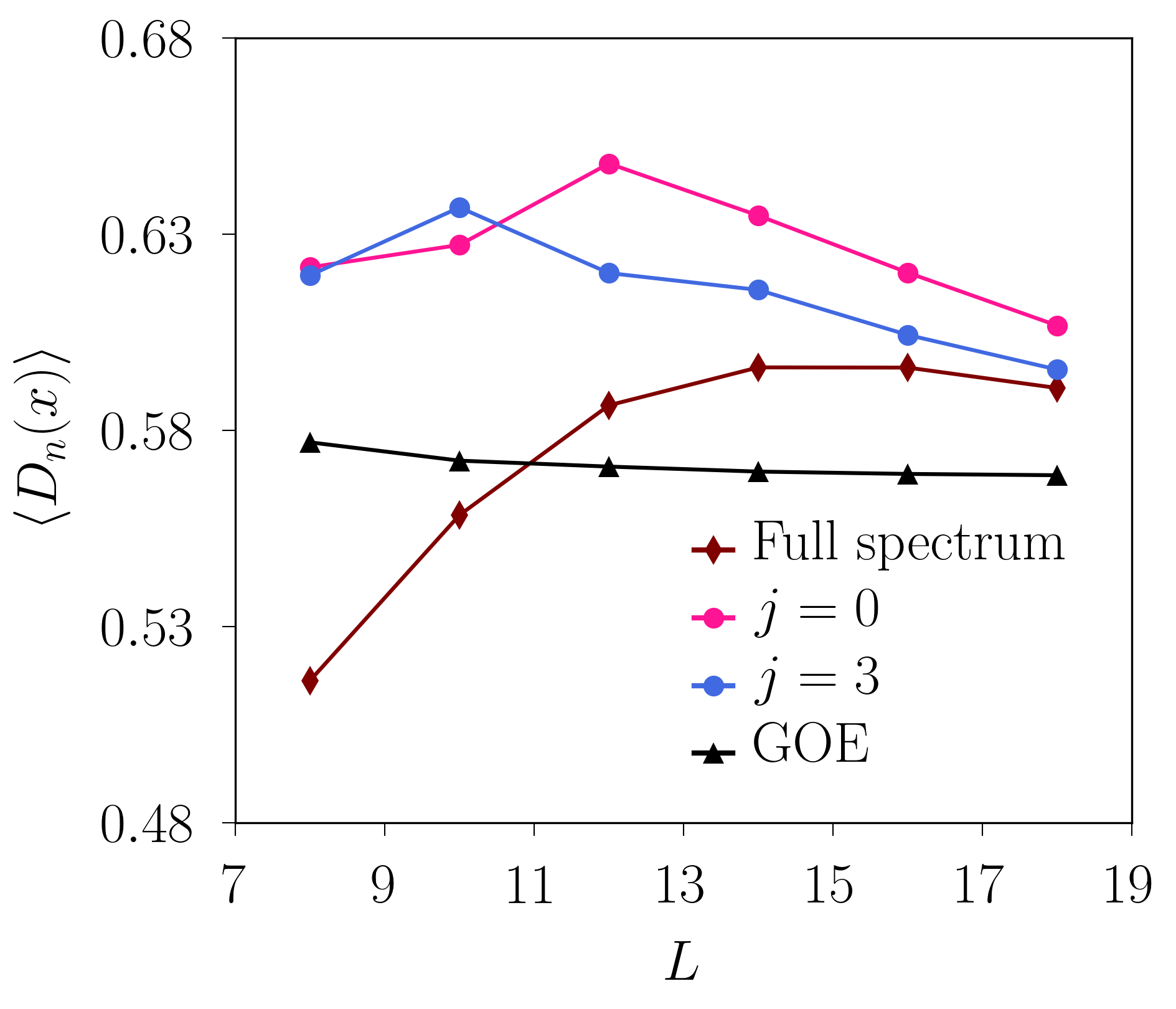}\put(-135,135){$(a)$}\put(-60,130){$x=1/2$}\put(-140,100){Chaotic Ising}
	\includegraphics[width=62mm]{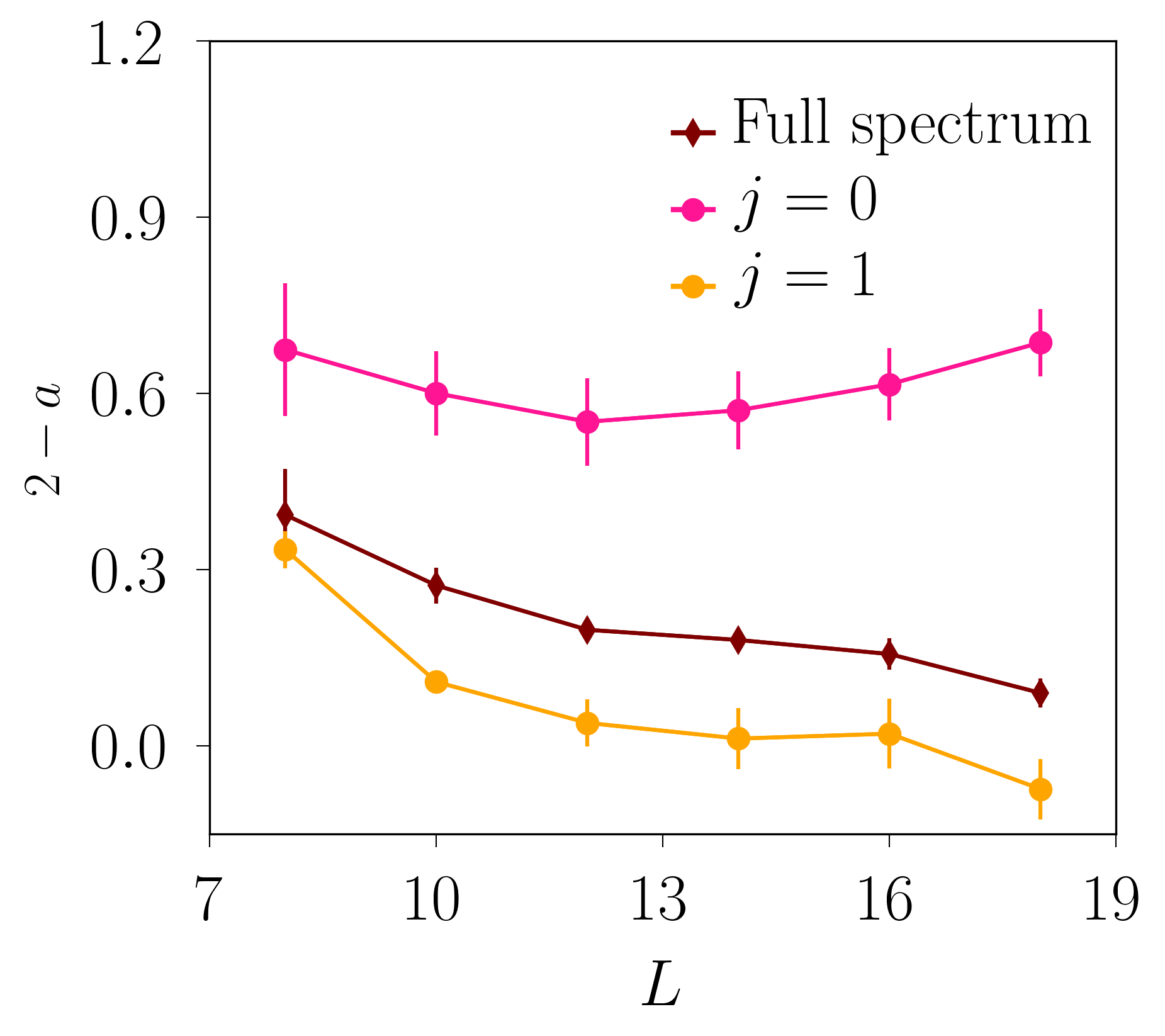}\put(-130,135){$(b)$}\put(-120,110){TFIM}
	\includegraphics[width=58mm]{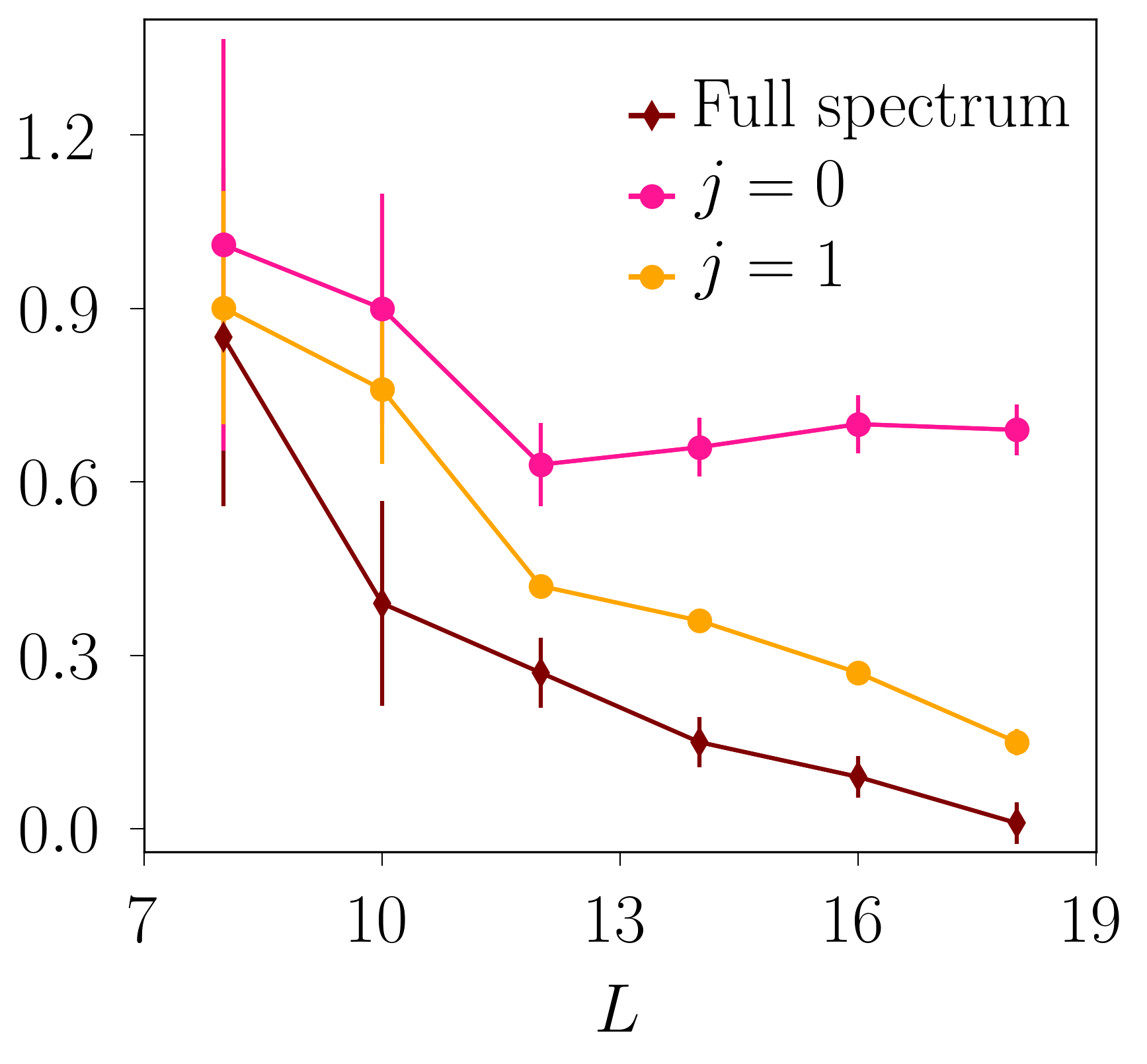}\put(-130,135){$(c)$}\put(-135,60){XXZ}
	\caption{Comparing full spectrum results with single symmetry sectors. $(a)$ The average of $D_{n}(x=1/2)$ as a function of system size for chaotic Ising model with  $h_x=(5+\sqrt{5})/8,h_z=(1+\sqrt{5})/4$. $(b),(c)$ The size-dependent behaviour of linear slope $a$ for TFIM with $h_x=(5+\sqrt{5})/8,h_z=0$ and XXZ chain with $\Delta=2.0$, respectively.}
	\label{fig:Ising_chaotic_full}
\end{figure*}

Therefore the energy fluctuations provides an upper bound on $||\frac{\partial\rho_n^A}{\partial\beta(\epsilon_n)}||$. Finally by replacing Eq.~\ref{eq:def_rho} and $\eta=(\Delta E)^2/(K_BT^2)$ in definition of $C$ we find that $C\leq 1/\Delta E$, therefore Eq.~\ref{eq:distance} takes the following form
\begin{equation}\label{eq:distance}
||\rho_{n+1}^A-\rho_n^A||\leq \frac{\Delta\epsilon_n}{\Delta E}.   
\end{equation}
In the thermodynamic limit, i.e., with the system size taken
to infinity, $\Delta\epsilon_n$ decays exponentially to zero by system sizes. The Eq.~\ref{eq:distance} insures similar behaviour for $||\rho_{n+1}^A-\rho_n^A||$.

\section{Full spectrum results and comparison with single symmetry sector}\label{app:symmetry sector}
In this section, we discuss the effect of Hamiltonian symmetries on $\langle D_{n}(x)\rangle$. The results presented in the main text are limited to a specific symmetry sector with momentum $K=2\pi/L$. Nevertheless, the statistics of eigenstates trace distances are independent of Hamiltonian symmetries. There are exceptions in the integrable regime, where the $K=0,\pi$ sectors do not exhibit the same universality as the other sectors and the entire spectrum. Nevertheless, the measure is still able to distinguish between integrable and chaotic regimes. 

We bring different examples to compare the full spectrum with single-sector results. We first consider the chaotic Ising model in Eq.~3. In Fig.~\ref{fig:Ising_chaotic_full}, the size-dependent behavior of $\langle D_{n}(x)\rangle$ with $x=1/2$ for the chaotic Ising model is shown. By growing $L$, we see the average distances for different symmetry sectors and the full spectrum approach to RMT value. For integrable models mentioned in the main text, $\langle D_{n}(x)\rangle$ follows a universal linear behavior for $x<0.5$ at fixed $L$. In Fig.~\ref{fig:Ising_chaotic_full}~$(a,b)$, the size dependency of the linear slope for TFIM and XXZ are shown, respectively.
\section{ Average trace distance in transverse field XY chain}\label{app:free_femions}
In this section, we consider the XY chain in the transverse field
\be
H = \sum_{l=1}^L \Big( \f{1+\g}{2} \s_j^x\s_{j+1}^x + \f{1-\g}{2} \s_j^y\s_{j+1}^y + h_z \s_j^z \Big)~,
\ee
where periodic boundary conditions are imposed such that $\s_{L+1}^{\a}=\s_{1}^{\a},\a=x,y,z$. The constant transverse field $h_z$ is also included in the Hamiltonian.

\begin{figure}[h]
  \includegraphics[width=0.32\textwidth]{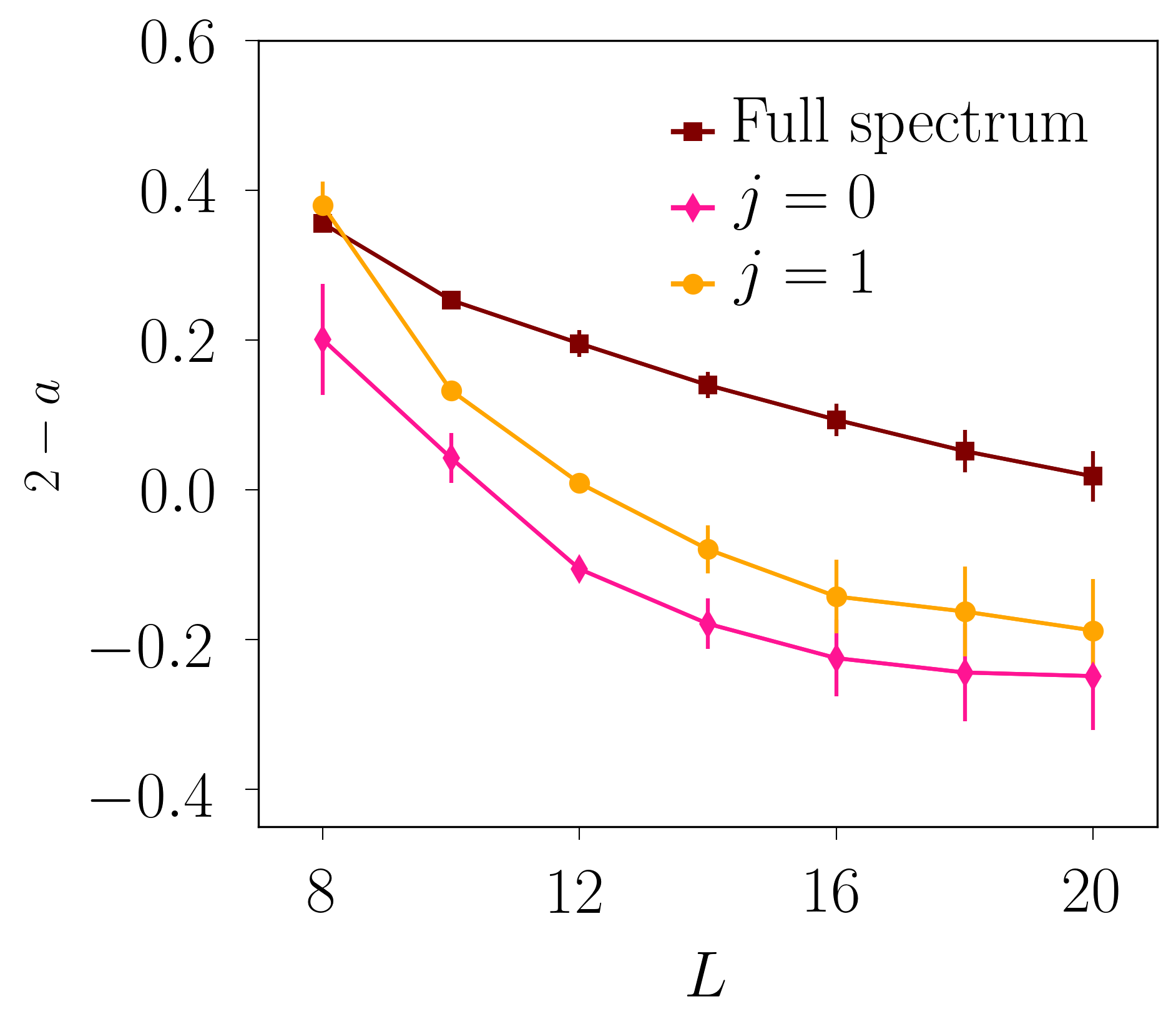}\put(-110,124){$(a)$}\put(-120,40){$\gamma=0.5,h_z=0.5$}
   \includegraphics[width=0.32\textwidth]{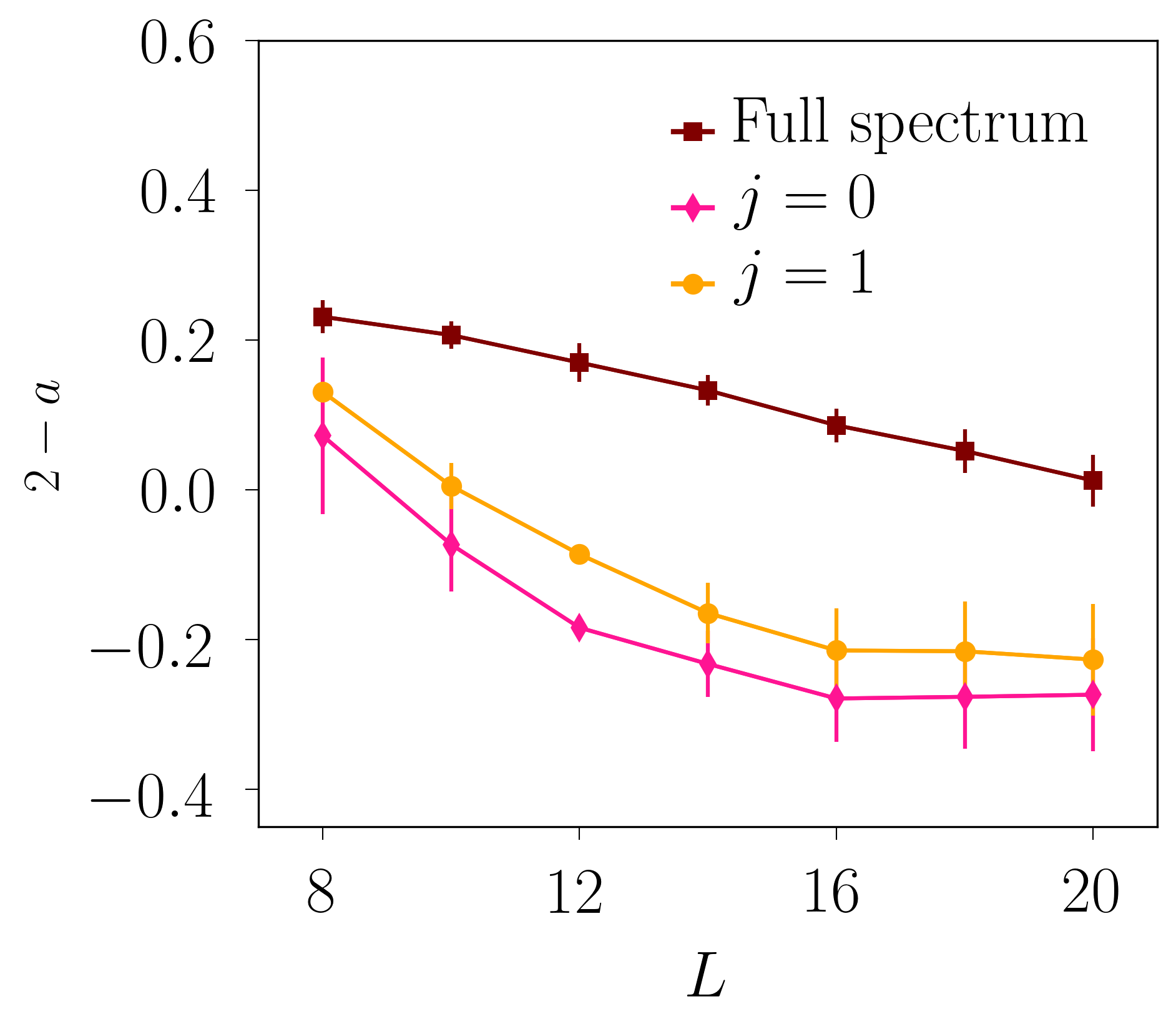}\put(-110,124){$(b)$}\put(-120,40){$\gamma=0.5,h_z=1.5$}
    \includegraphics[width=0.32\textwidth]{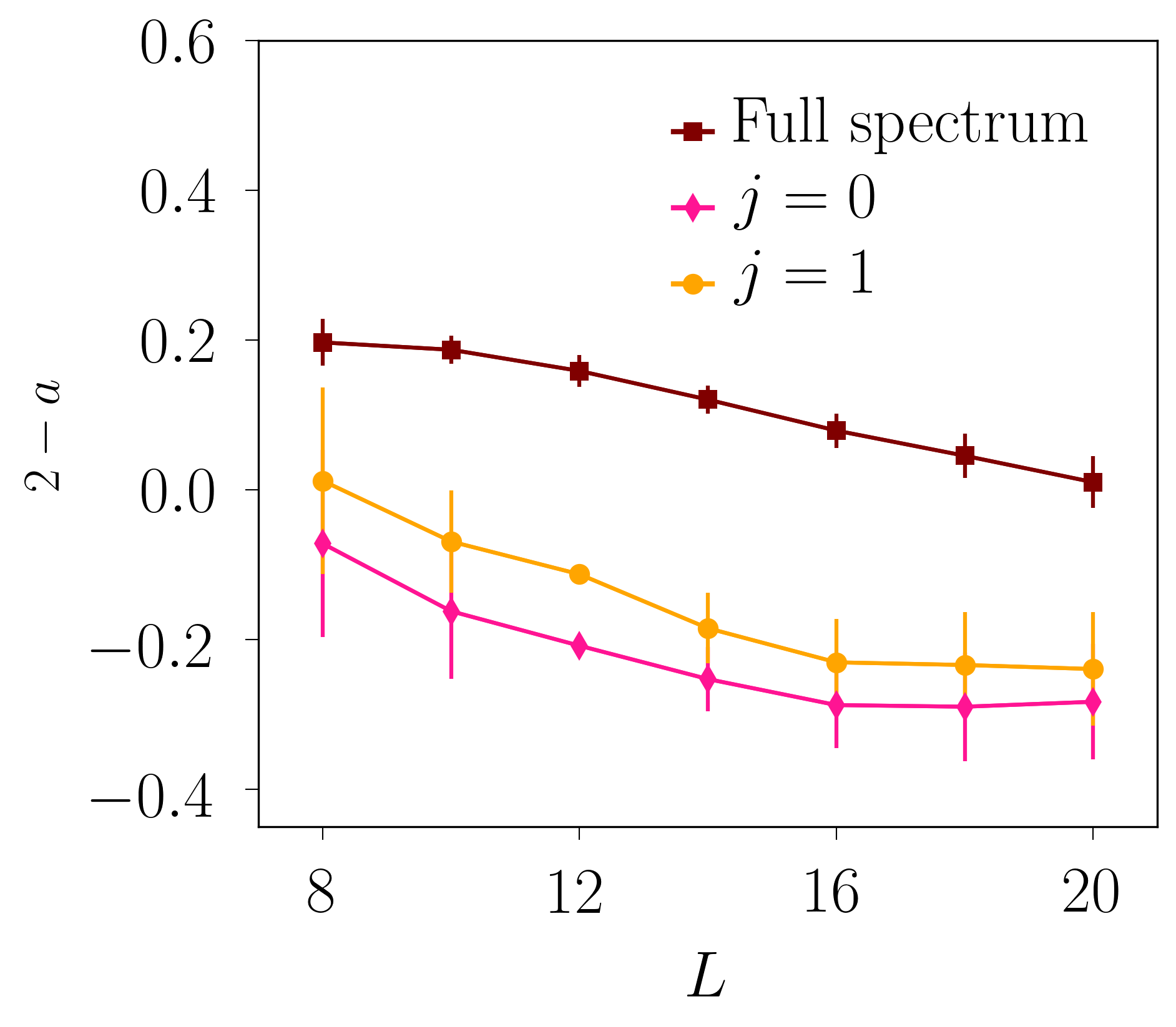}\put(-110,124){$(c)$}\put(-120,40){$\gamma=1.5,h_z=2.0$}\\
  \caption{Slope of the average trace distance in the XY chain in the transverse field with different parameters from integrability techniques. In large $L$ limit, for the whole spectrum the slope $a \to 2$, while for a single momentum block the slope $a$ approaches a value that is larger than 2.}
  \label{XYintslope}
\end{figure}

Using integrability techniques, we determine the slope of the average trace distance for both the entire spectrum and a block of fixed momentum, as shown in Fig.~\ref{XYintslope}. When calculating for the entire spectrum, we first sort the states based on energy and then sort the energy degenerate states based on momentum. In the limit of large $L$, the slope $a \to 2$ for the entire spectrum, which is the same as in the first two panels of Fig~3 in the main text. However, for a single momentum block, the slope $a$ approaches a value greater than 2.

\begin{figure}[h]
  \centering
  \includegraphics[width=0.35\textwidth]{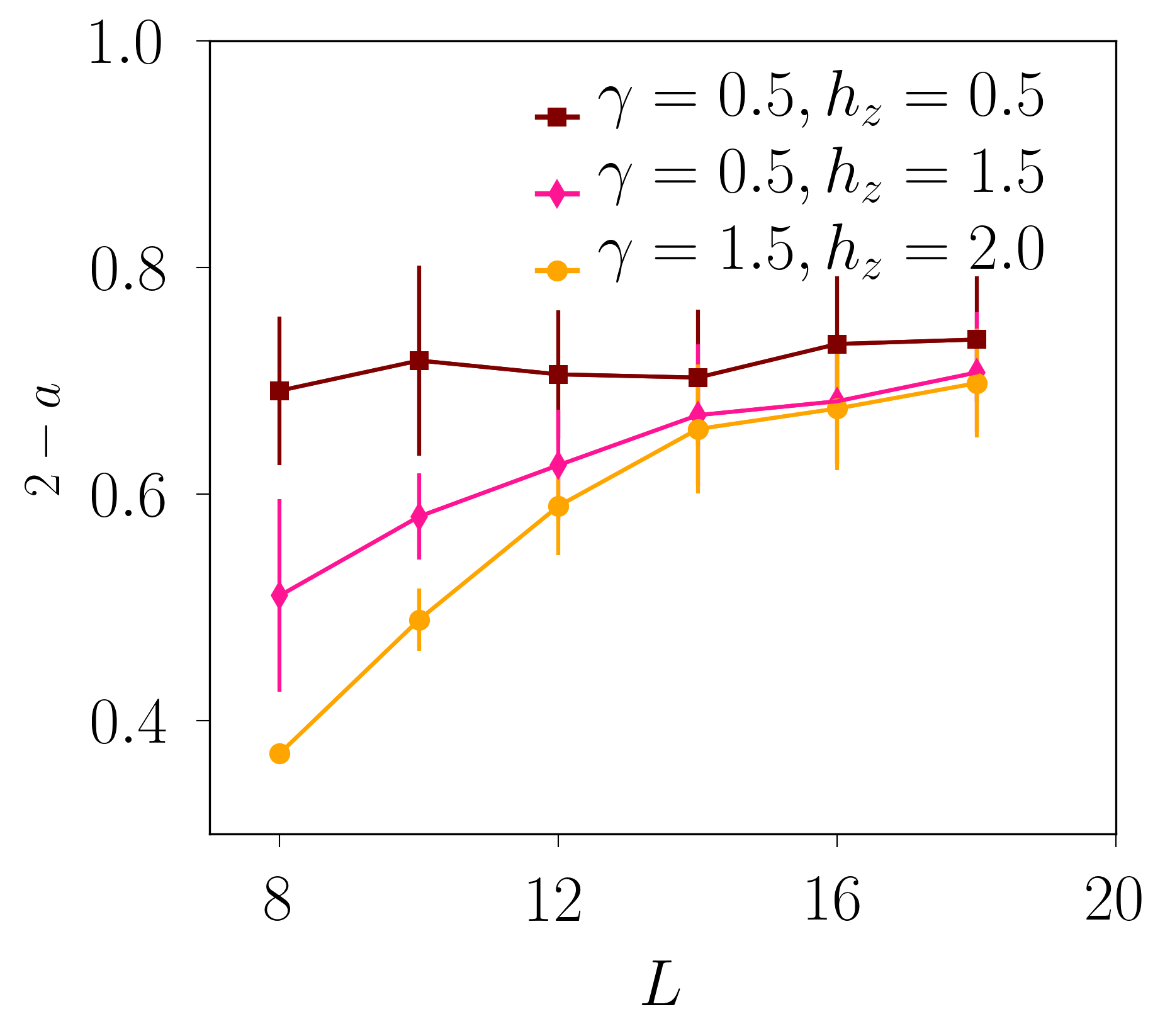}\\
  \caption{Slope of the average trace distance in the XY chain in the transverse field with different parameters from exact diagonalization. In large $L$ limit, for momentum block $j=0$ the slope $a$ approaches a value that is smaller than 2.}
  \label{XYedslope}
\end{figure}

We also calculate the slope of average trace distance for both the whole spectrum and the block with fixed momentum from exact diagonalization, as depicted in Fig.~\ref{XYedslope}.
In large $L$ limit, for the whole spectrum and a single block with momentum $j\neq0$ and $j\neq\f{L}{2}$, the slope $a \to 2$. But for a single block with $j=0$ or $j=\f{L}{2}$, the slope $a$ approaches a value that is smaller than 2, as we show in Fig.~\ref{XYedslope}.
This could be attributed to the fact that the majority of the states in the $j=0$ and $j=\f{L}{2}$ blocks are degenerate. The average trace distance is not well-defined for the $j=0$ and $j=\f{L}{2}$ blocks, and so we obtain different results from different methods.

\section{Quantum kicked top model}\label{app:QKT}
\begin{figure*}
    \includegraphics[width=48mm]{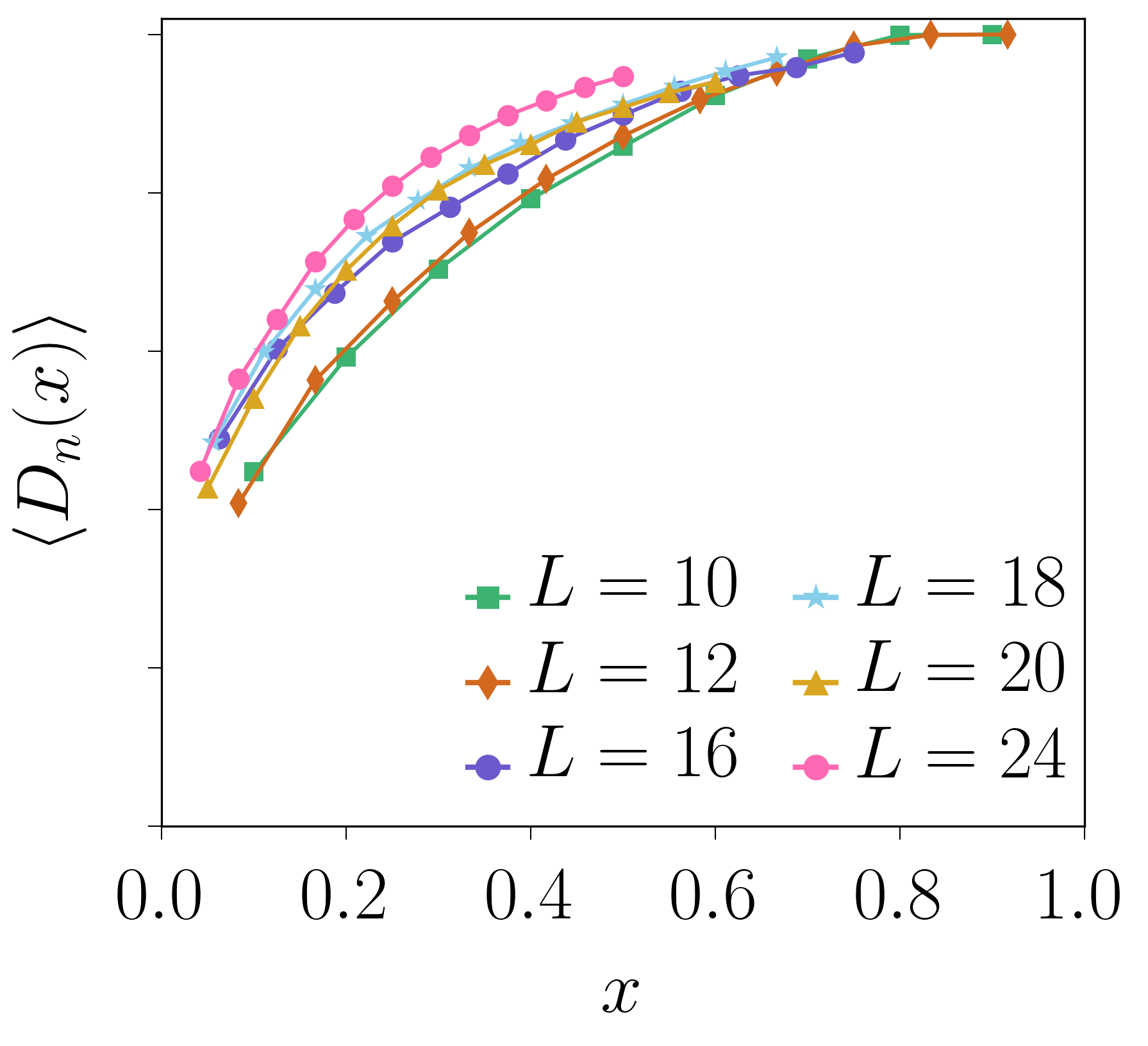}\put(-110,105){$(a)$}\put(-50,85){$\kappa=0.4$}
	\includegraphics[width=43.5mm]{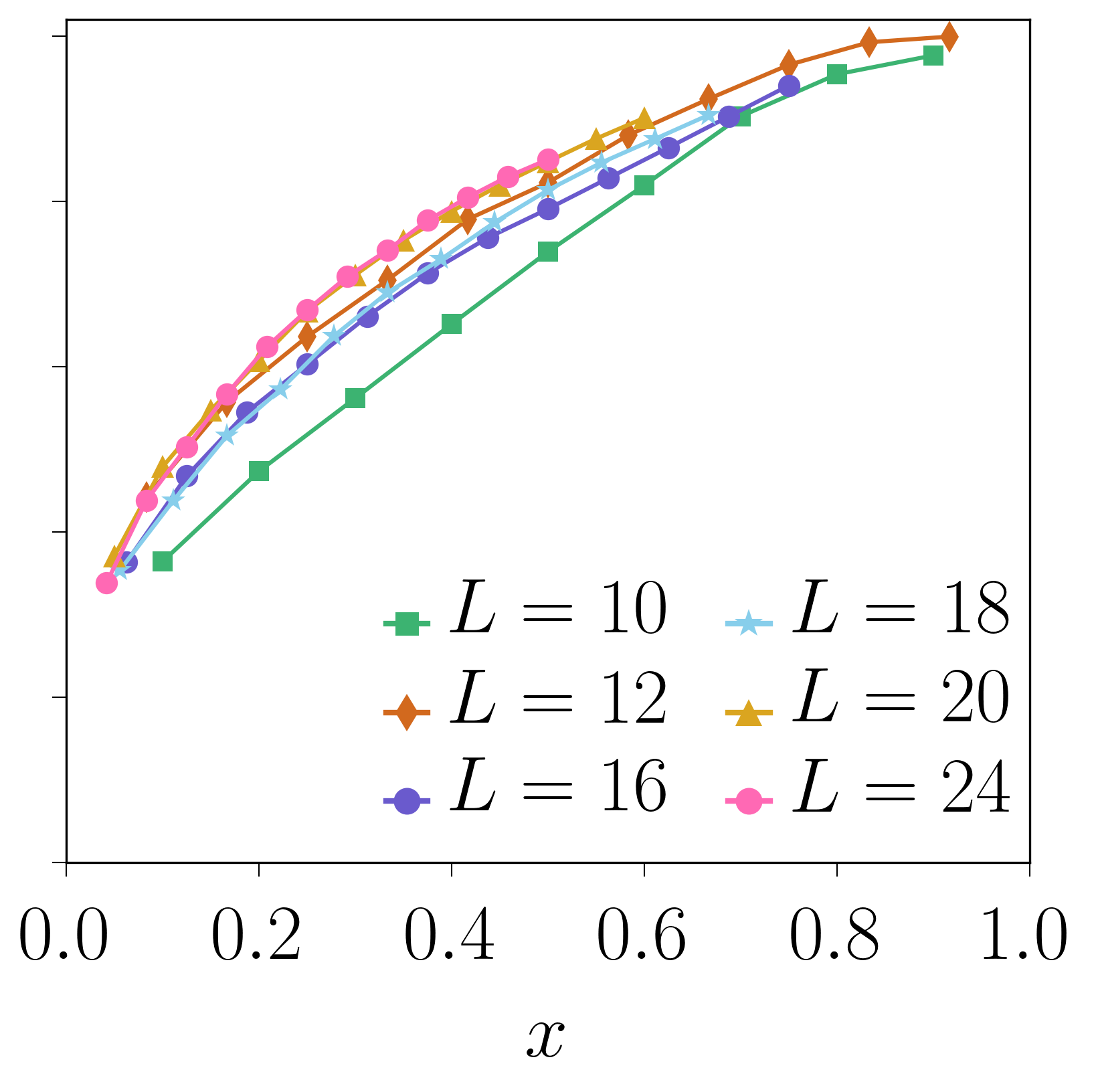}\put(-110,105){$(b)$}\put(-50,85){$\kappa=1.7$}
	\includegraphics[width=43.5mm]{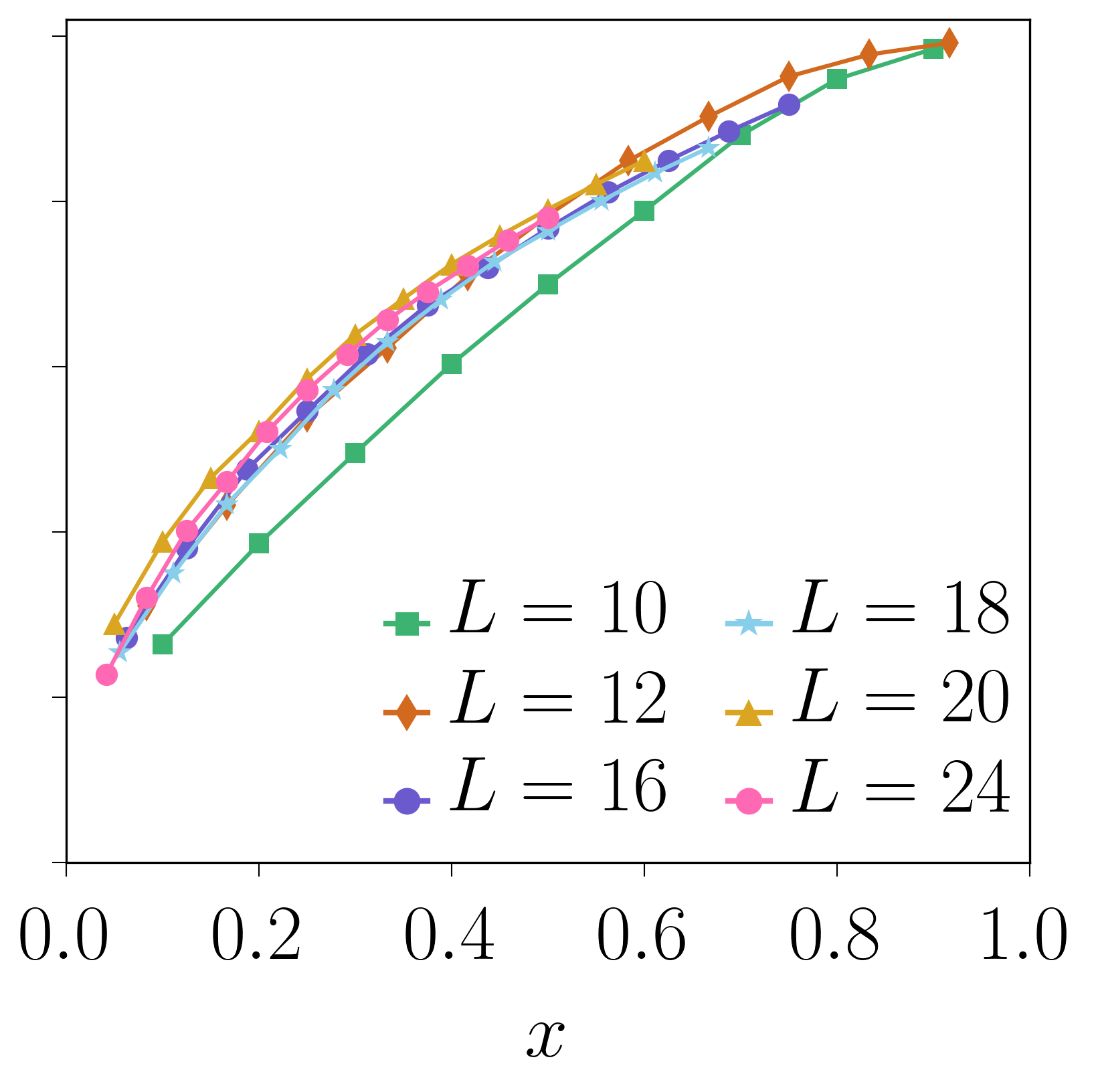}\put(-110,105){$(c)$}\put(-50,85){$\kappa=3.0$}
	\includegraphics[width=43.5mm]{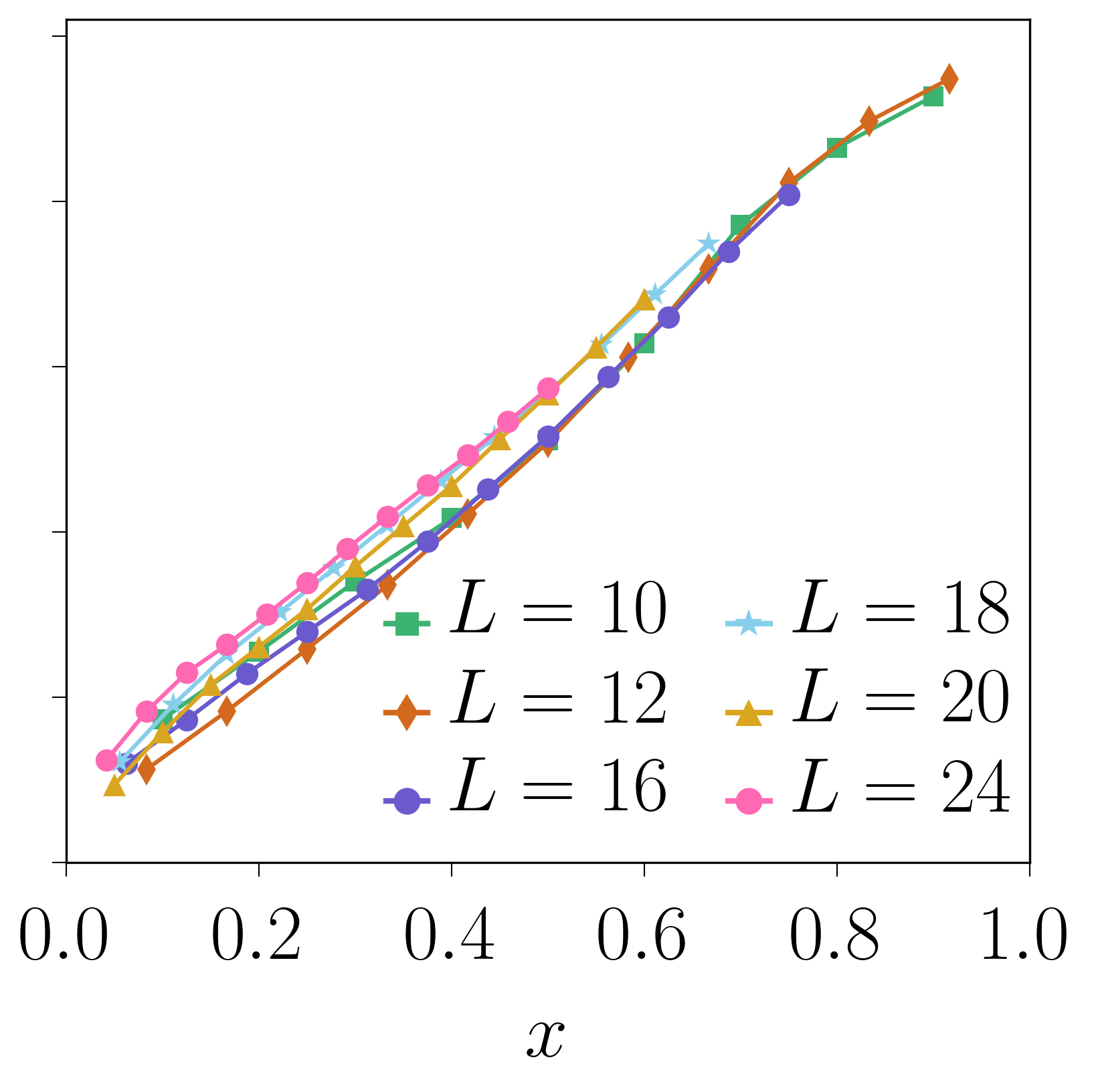}\put(-110,105){$(d)$}\put(-110,85){$\kappa=7.0$}
	\caption{{\bf Quantum kicked top model}: $\langle D_{n}(x) \rangle$ as a function of $x=L_A/L$. Here $\tau=1$.}
	\label{fig:QKT}
\end{figure*}

In this supplementary material, we provide more comprehensive examples of $\langle D_n(x)\rangle$ for the quantum kicked top model. As previously stated in the main text, this model can be classified as an integrable model in the many-body context. However, it is worth noting that there exists a region of the parameter $\kappa$ where the behavior of $\langle D_n(x)\rangle$ is consistent with that of a MBL. The numerical results for a broad range of the parameter $\kappa$ are presented in Fig.~\ref{fig:QKT}.

\section{Eigenstate-to-eigenstate fluctuations of local observable}\label{app:local_obs}
\begin{figure*}
    \includegraphics[width=48mm]{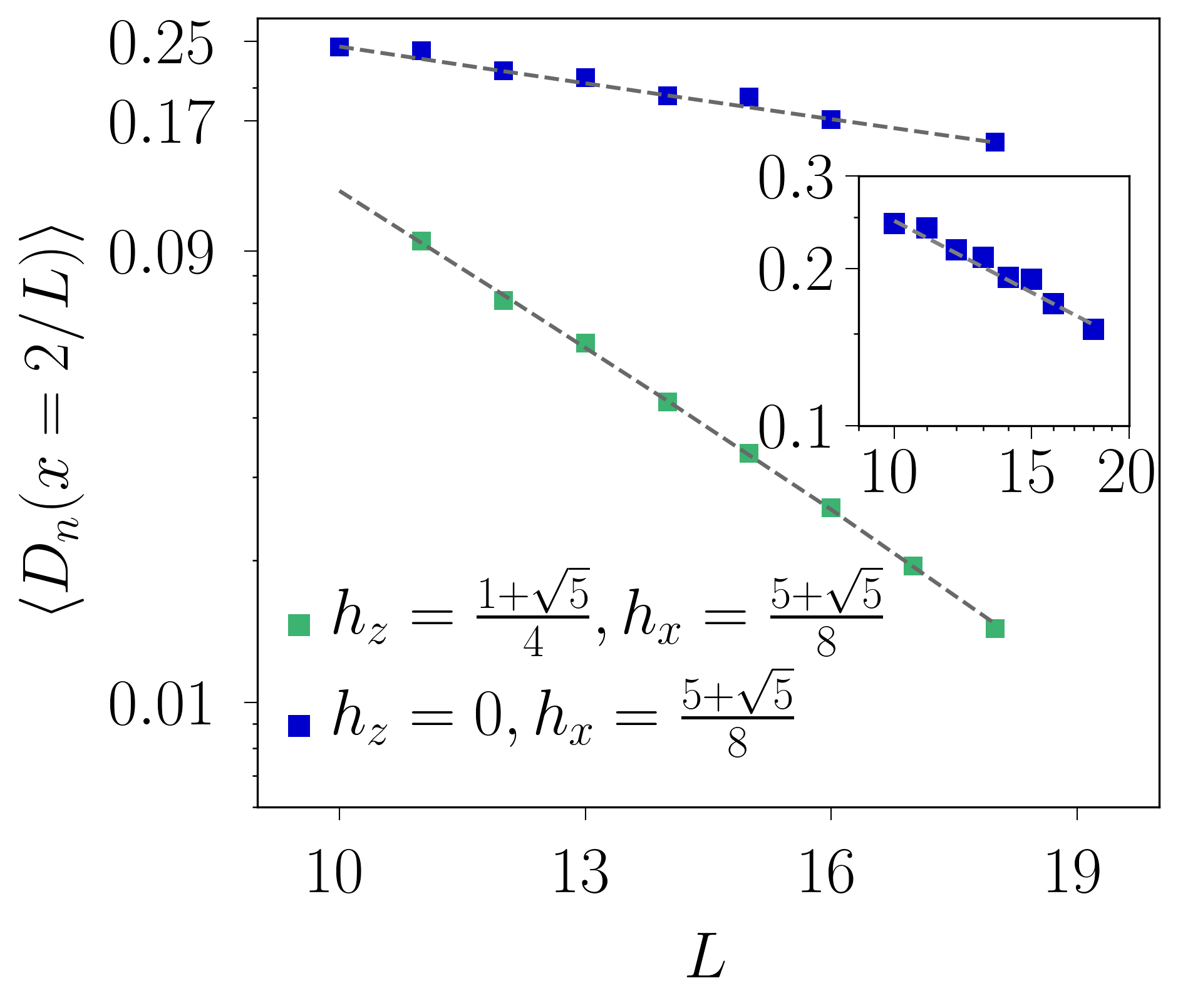}\put(-20,105){$(a)$}
	\includegraphics[width=49mm]{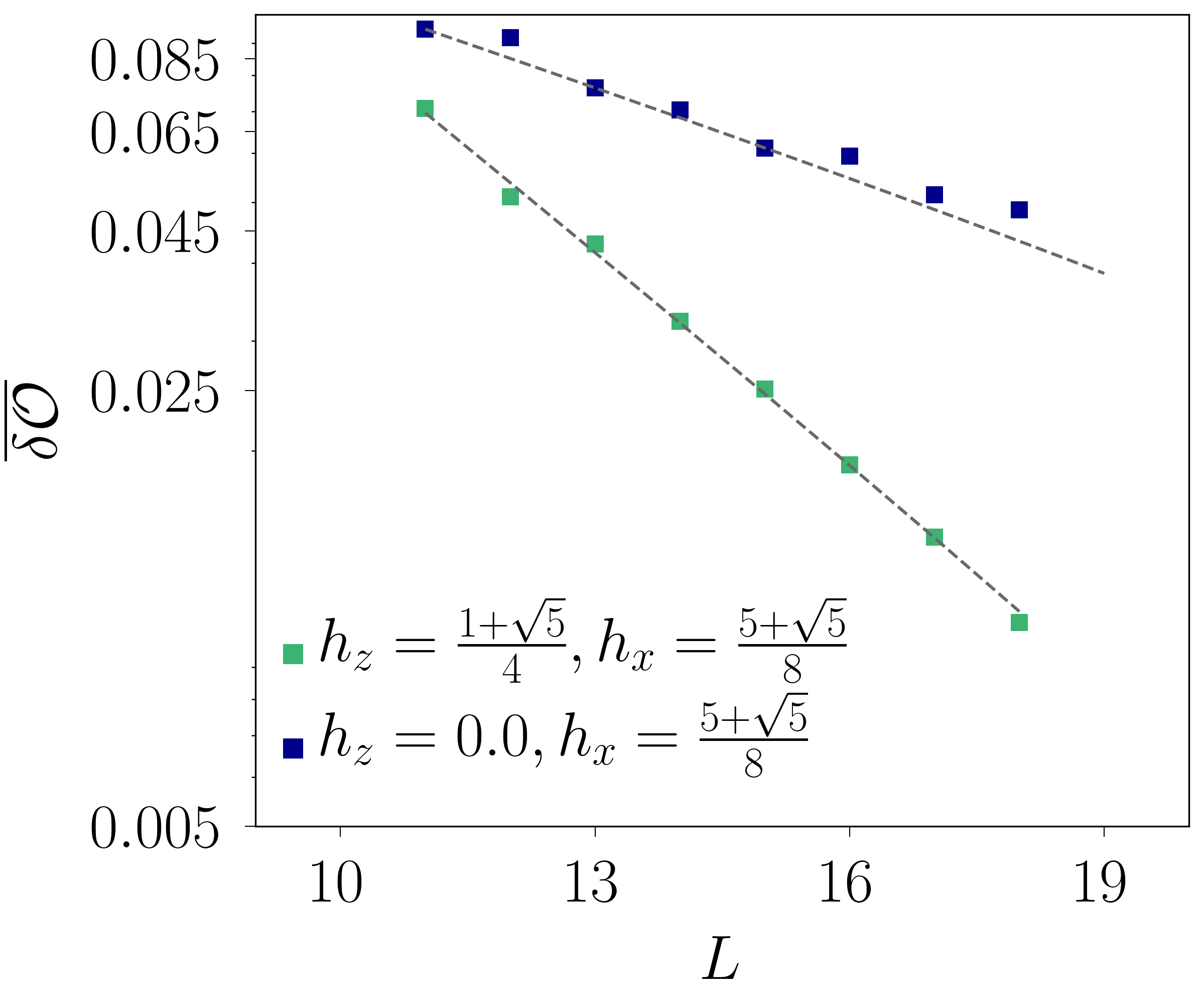}\put(-20,105){$(b)$}
	\caption{{\bf Ising model}: Size dependent behaviour of average $(a)$ trace distances $\langle D_{n}(x)\rangle$, and $(b)$ eigenstate-to-eigenstate fluctuations of observable $\overline{\delta{\cal O}}$ with ${\cal O}=\sigma_1^x\sigma_2^x$. The average is taken over the symmetry-sector at momentum $K=2\pi/L$.} 
	\label{fig:obs}
\end{figure*}

In the main text, Eq.~(4) defines trace distances that serve as an upper bound for all conceivable local observables, extending even to certain non-local ones such as entanglement entropy. This relationship impinges on the broader understanding of quantum systems, where trace distances provide a significant theoretical tool. In the subsequent analysis, we demonstrate the consistency of our results of trace distances when examined in conjunction with two-site observables. Fig.\ref{fig:obs} illustrates the system size-dependent behavior of $\langle D_n(x)\rangle$ and $\overline{\delta{\cal O}}=\sum_{n\in \{K=2\pi/L\}} \big(\bra{\psi_{n+1}}{\cal O}\ket{\psi_{n+1}}-\bra{\psi_{n}}{\cal O}\ket{\psi_{n}}\big)$ with ${\cal O}=\sigma_1^x\sigma_2^x$, specifically in a fixed $L_A=2$. According to our numerical analysis for the quantum Ising model studied in the main text, we find that in the chaotic regime, defined by $h_x=(5+\sqrt{5})/8,h_z=(1+\sqrt{5})/4$, the average of trace distances alongside the selected local observable illustrates a rapid exponential decline as a function of $L$. Within the integrable regime and concomitant with a more gradual decay as the system size increases, we note pronounced finite-size effects. These effects preclude us from definitively asserting that the decay in this regime conforms to a power law. This result agrees with the recent work of Ref.~\cite{Rigol:2019}, which thoroughly investigates the system size-dependent behavior of local observables in chaotic and integrable Hamiltonians. This finding underscores the importance of considering an extensive range of $x$, as undertaken in the present study, to effectively differentiate between many-body integrable and chaotic phases.

\hfill\\
\newpage
\providecommand{\href}[2]{#2}\begingroup\raggedright\endgroup


\begin{thebibliography}{10}
%
\bibitem{Gutzwiller:2013}
M. C. Gutzwiller, \textit{{Chaos in classical and quantum mechanics}}, \href{https://link.springer.com/book/10.1007/978-1-4612-0983-6}{Springer Science \& Business Media. {\bfseries 1}(2013)}.
%
\bibitem{Babelon:2003}
O. Babelon, D. Bernard, and M. Talon, \textit{{ Introduction to classical integrable systems}}, \href{}{Cambridge Univ. Press (2003)}.
%
\bibitem{Caux:2011}
J. S. Caux, and J. Mossel, \textit{{Remarks on the notion of quantum integrability}}, \href{https://doi.org/10.1088/1742-5468/2011/02/P02023}{J. Stat. Mech.: Theory Exp. {\bfseries 2011}, P02023 (2011)}[\href{https://doi.org/10.48550/arXiv.1012.3587}{{\ttfamily  	arXiv:1012.3587}}].
%
\bibitem{Tabor:1977}
M. V. Berry and M. Tabor, \textit{{Level clustering in the regular spectrum}}, \href{https://doi.org/10.1098/rspa.1977.0140}{Proc. R. Soc. Lond A. {\bfseries 356}, 375 (1977)}
%
\bibitem{Bohigas:1984}
O. Bohigas, M. J. Giannoni, and C. Schmit, \textit{{Characterization of chaotic quantum spectra and universality of level fluctuation laws}}, \href{https://doi.org/10.1103/PhysRevLett.52.1}{Phys. Rev. Lett. {\bfseries 52}, 1 (1984)}.
%
\bibitem{Finkel:2005}
F. Finkel, and A. Gonz{\'a}lez-L{\'o}pez, \textit{{Global properties of the spectrum of the Haldane-Shastry spin chain}}, \href{https://doi.org/10.1103/PhysRevB.72.174411}{Phys. Rev.
B. {\bfseries 72}, 174411 (2005)}[\href{https://doi.org/10.48550/arXiv.cond-mat/0509032}{{\ttfamily  	 	 	arXiv:cond-mat/0509032}}].
%
%

\bibitem{Sieberer:2019}
L. M. Sieberer, T. Olsacher, A. Elben, M. Heyl, P. Hauke, F. Haake, and P. Zoller, \textit{{Digital quantum simulation, Trotter errors, and quantum chaos of the kicked top}}, \href{ https://doi.org/10.1038/s41534-019-0192-5}{Npj Quantum Inf. {\bfseries 5}, 1-11 (2019)}[\href{ 	
https://doi.org/10.48550/arXiv.1812.05876}{{\ttfamily arXiv:1812.05876}}].
%
\bibitem{Dodonov:2000}
V.V. Dodonov, O.V. Man'Ko, V.I. Man'Ko, and  A. W{\"u}nsche, \textit{{Hilbert-Schmidt distance and non-classicality of states in quantum optics}}, \href{https://doi.org/10.1080/09500340008233385}{J. Mod. Opt. {\bfseries 47}, 633-654 (2000)}.
%
\bibitem{Gilchrist:2005}
A. Gilchrist, N. K. Langford, and M. A. Nielsen, \textit{{Distance measures to compare real and ideal quantum processes}}, \href{https://doi.org/10.1103/PhysRevA.71.062310}{Phys. Rev. A. {\bfseries 71}, 062310 (2005)}, [\href{ 	
	https://doi.org/10.48550/arXiv.quant-ph/0408063}{{\ttfamily arXiv:quant-ph/0408063}}].
%
\bibitem{Nielsen:2009}
M. A. Nielsen, and I. L. Chuang, \textit{{Quantum computation and quantum information}}, \href{http://mmrc.amss.cas.cn/tlb/201702/W020170224608149940643.pdf}{Cambridge University Press, Cambridge, UK, ISBN 9781107002173 (2009)}.
%
\bibitem{Rajabpour:2020}
J. Zhang, P. Calabrese, M. Dalmonte, and M. A. Rajabpour, \textit{{Lattice Bisognano-Wichmann modular Hamiltonian in critical quantum spin chains}}, \href{https://doi.org/10.21468/SciPostPhysCore.2.2.007}{SciPost Phys. Core. {\bfseries 2}, 007 (2020)}, [\href{https://doi.org/10.48550/arXiv.2003.00315}{{\ttfamily arXiv:2003.00315}}].
%
\bibitem{Rigol:2008}
M. Rigol, V. Dunjko, and M. Olshanii, \textit{{Thermalization and its mechanism for generic isolated quantum systems}}, \href{http://www.nature.com/doifinder/10.1038/nature06838}{Nature. {\bfseries 452}, 854-858 (2008)}, [\href{ 	
	https://doi.org/10.48550/arXiv.0708.1324}{{\ttfamily arXiv:0708.1324}}].
%
\bibitem{Deutsch:2018}
J. M. Deutsch, \textit{{Eigenstate thermalization hypothesis}}, \href{https://doi.org/10.1088/1361-6633/aac9f1}{Rep. Prog. Phys. {\bfseries 81}, 082001 (2018)}, [\href{ 	
	https://doi.org/10.48550/arXiv.1805.01616}{{\ttfamily arXiv:1805.01616 }}].
%
\bibitem{Dymarsky:2016ntg}
A.~Dymarsky, N.~Lashkari and H.~Liu, \textit{{Subsystem eigenstate
  thermalization hypothesis}},
  \href{http://dx.doi.org/10.1103/PhysRevE.97.012140}{Phys. Rev. E {\bfseries
  97}, 012140 (2018)}, [\href{https://arxiv.org/abs/1611.08764}{{\ttfamily
  arXiv:1611.08764}}].
%
%
\bibitem{Chehade:2019}
S. S. Chehade and A. Vershynina, \textit{{Quantum entropies}}, \href{http://www.scholarpedia.org/article/Quantum_entropies}{Scholarpedia. {\bfseries 14}, 53131 (2019)}.
%
\bibitem{Fannes:1973}
M. Fannes, \textit{{A continuity property of the entropy density for spin lattice systems}}, \href{https://link.springer.com/content/pdf/10.1007/BF01646490.pdf}{Commun. Math. Phys. {\bfseries 31}, 291 (1973)}.
%
\bibitem{Audenaert:2007}
K. M. R. Audenaert, \textit{{A sharp continuity estimate for the von Neumann entropy}}, \href{ 	
	%https://doi.org/10.1088/1751-8113/40/28/S18}{J. Phys. A: Math. Theor. {\bfseries 40}, 8127 (2007)}.
%
\bibitem{Vidmar:2017}
L. Vidmar, and M. Rigol, \textit{{Entanglement Entropy of Eigenstates of Quantum Chaotic Hamiltonians}}, \href{ https://doi.org/10.1103/PhysRevLett.119.220603}{Phys. Rev. L. {\bfseries 119}, 220603 (2017)}[\href{ 	
https://doi.org/10.48550/arXiv.1708.08453}{{\ttfamily arXiv:1708.08453}}].
%
\bibitem{Hackl:2017}
L. Vidmar, L. Hackl, E. Bianchi, and M. Rigol, \textit{{Entanglement Entropy of Eigenstates of Quadratic Fermionic Hamiltonians}}, \href{https://doi.org/10.1103/PhysRevLett.119.020601}{Phys. Rev. L. {\bfseries 119}, 020601 (2017)}[\href{ 	
https://doi.org/10.48550/arXiv.1703.02979}{{\ttfamily arXiv:1703.02979}}].
%
\bibitem{Blond:2019}
T. LeBlond, K. Mallayya, L. Vidmar, and M. Rigol, \textit{{Entanglement and matrix elements of observables in interacting integrable systems}}, \href{ 	
https://doi.org/10.1103/PhysRevE.100.062134}{Phys. Rev. E. {\bfseries 100}, 062134 (2019)}[\href{https://doi.org/10.48550/arXiv.1909.09654}{{\ttfamily arXiv:1909.09654}}].
%
%
\bibitem{Supplement}
See Supplemental Material at \href{ https://doi.org/...}{https://doi.org/...} for the detailed calculations of ransom matrix theory, analytical calculations using eigenstate thermalization hypothesis, robustness of trace distance measure to Hamiltonian symmetries.
%
\bibitem{Huse:2013}
H. Kim and D. A. Huse, \textit{{Ballistic Spreading of Entanglement in a Diffusive Nonintegrable System}}, \href{https://doi.org/10.1103/PhysRevLett.111.127205}{ Phys. Rev. Lett. {\bfseries 111}, 127205 (2013)}, [\href{https://doi.org/10.48550/arXiv.1306.4306}{{\ttfamily arXiv:1306.4306v2}}].
%
\bibitem{Miranda:2023}
J. T. de Miranda, and T. Micklitz, \textit{{Subsystem trace-distances of two random states}}, \href{ 	
https://doi.org/10.1088/1751-8121/acc770}{ J. Phys. A: Math. Theor. {\bfseries 56}, 175301 (2023)}, [\href{ 	
https://doi.org/10.48550/arXiv.2210.03213}{{\ttfamily arXiv:2210.03213}}].
%
\bibitem[]{footnote1}
We note that due to the large number of degeneracies in these sectors there is an ambiguity in ordering the states, which leads to different slops depending on the method that one uses, see the Supplemental Material \cite{Supplement}

\bibitem{Lieb:1961fr}
E. Lieb, T. Schultz and D. Mattis, \textit{{Two soluble models of an
  antiferromagnetic chain}},
  \href{http://dx.doi.org/10.1016/0003-4916(61)90115-4}{Annals Phys. {\bfseries
  16}, 407 (1961)}.

\bibitem{Katsura:1962hqz}
S.~Katsura, \textit{{Statistical mechanics of the anisotropic linear Heisenberg
  model}}, \href{http://dx.doi.org/10.1103/PhysRev.127.1508}{{Phys. Rev.}
  {\bfseries 127}, 1508 (1962)}.

\bibitem{Pfeuty:1970ayt}
P.~Pfeuty, \textit{{The one-dimensional Ising model with a transverse field}},
  \href{http://dx.doi.org/10.1016/0003-4916(70)90270-8}{Annals Phys. {\bfseries
  57}, 79 (1970)}.


\bibitem{Zhang:2019wqo}
J. Zhang, P. Ruggiero, and P.~Calabrese, \textit{{Subsystem trace distance in
  quantum field theory}},
  \href{http://dx.doi.org/10.1103/PhysRevLett.122.141602}{Phys. Rev. Lett.
  {\bfseries 122}, 141602 (2019)},
  [\href{https://arxiv.org/abs/1901.10993}{{\ttfamily arXiv:1901.10993}}].

\bibitem{Zhang:2019itb}
J. Zhang, P. Ruggiero, and P. Calabrese, \textit{{Subsystem trace distance in
  low-lying states of $(1+1)$-dimensional conformal field theories}},
  \href{http://dx.doi.org/10.1007/JHEP10(2019)181}{JHEP {\bfseries 10} (2019)
  181}, [\href{https://arxiv.org/abs/1907.04332}{{\ttfamily
  arXiv:1907.04332}}].

\bibitem{Zhang:2022tgu}
J.~Zhang and M.~A. Rajabpour, \textit{{Subsystem distances between
  quasiparticle excited states}},
  \href{http://dx.doi.org/10.1007/JHEP07(2022)119}{JHEP {\bfseries 07} (2022)
  119}, [\href{https://arxiv.org/abs/2202.11448}{{\ttfamily
  arXiv:2202.11448}}].
%
%
%
%
\bibitem{Marco:2008}
M. {\v{Z}}nidari{\v{c}}, T. Prosen, and P. Prelov{\v{s}}ek, \textit{{Many-body localization in the Heisenberg XXZ magnet in a random field}}, \href{https://doi.org/10.1103/PhysRevB.77.064426}{ Phys. Rev. B. {\bfseries 77}, 064426 (2008)}, [\href{https://doi.org/10.48550/arXiv.0706.2539 }{{\ttfamily arXiv:0706.2539}}].
%
\bibitem{Huse:2014}
D. A. Huse, R. Nandkishore, and V. Oganesyan, \textit{{Phenomenology of fully many-body-localized systems}}, \href{ 	
https://doi.org/10.1103/PhysRevB.90.174202}{ Phys. Rev. B. {\bfseries 90}, 174202 (2014)}, [\href{https://doi.org/10.48550/arXiv.1408.4297 }{{\ttfamily arXiv:1408.4297 }}].
%
\bibitem{haake:1987}
F. Haake, M. Ku{\'s}, and R. Scharf, \textit{{Classical and quantum chaos for a kicked top}}, \href{ 	
https://link.springer.com/content/pdf/10.1007/BF01303727.pdf}{ Zeitschrift f{\"u}r Physik B Condensed Matter. {\bfseries 65}, 381-395 (1987)}.
%
%






\bibitem{Wang:2004}
X. Wang, S. Ghose, B. C. Sanders, and B. Hu, \textit{{Entanglement as a signature of quantum chaos}}, \href{
https://doi.org/10.1103/PhysRevE.70.016217}{ Phys. Rev. E. {\bfseries 70}, 016217 (2004)}, [\href{ 	
https://doi.org/10.48550/arXiv.quant-ph/0312047}{{\ttfamily  	arXiv:quant-ph/0312047}}].
%
\bibitem{Milburn:1999}
G. J. Milburn, \textit{{Simulating nonlinear spin models in an ion trap,}}[\href{ 	
https://doi.org/10.48550/arXiv.quant-ph/9908037}{{\ttfamily arXiv:quant-ph/9908037}}].
%
\bibitem{Dogra:2019}
S. Dogra, V. Madhok, and A. Lakshminarayan, \textit{{Quantum signatures of chaos, thermalization, and tunneling in the exactly solvable few-body kicked top}}, \href{ 	
https://doi.org/10.1103/PhysRevE.99.062217}{ Phys. Rev. E. {\bfseries 99}, 062217 (2019)}, [\href{https://doi.org/10.48550/arXiv.1902.10769 }{{\ttfamily arXiv:1902.10769}}].
%
\bibitem{Alicki:1996}
R. Alicki, D. Makowiec, and W. Miklaszewski, \textit{{Quantum Chaos in Terms of Entropy for a Periodically Kicked Top}}, \href{https://doi.org/10.1103/PhysRevLett.77.838}{ Phys. Rev. L. {\bfseries 77}, 838 (1996)}.
%
\bibitem{Ghose:2008}
 S. Ghose, R. Stock, P. Jessen, R. Lal, and A. Silberfarb, \textit{{Chaos, entanglement, and decoherence in the quantum kicked top}}, \href{https://doi.org/10.1103/PhysRevA.78.042318}{ Phys. Rev. A. {\bfseries 78}, 042318 (2008)}, [\href{https://doi.org/10.48550/arXiv.0805.1264}{{\ttfamily  	 	 	arXiv:0805.1264}}].
%
\bibitem{Wang:2021}
 Q. Wang, and M. Robnik, \textit{{Multifractality in quasienergy space of coherent states as a signature of quantum chaos}}, \href{https://doi.org/10.3390/e23101347}{ Entropy. {\bfseries 23}, 1347 (2021)}, [\href{https://doi.org/10.48550/arXiv.2110.10509}{{\ttfamily  	arXiv:2110.10509}}].
%
\bibitem{Suntajs:2020}
 J. {\v{S}}untajs, J. Bon{\v{c}}a, T. Prosen, and L. Vidmar, \textit{{Quantum chaos challenges many-body localization}}, \href{ 	
https://doi.org/10.1103/PhysRevE.102.062144}{ Phys. Rev. E. {\bfseries 102}, 062144 (2020)}, [\href{https://doi.org/10.48550/arXiv.1905.06345}{{\ttfamily arXiv:1905.06345}}].
%
\bibitem{Abanin:2021}
 D.A. Abanin, J.H. Bardarson, G. De Tomasi, S. Gopalakrishnan, V. Khemani, S.A. Parameswaran, F. Pollmann, A.C. Potter, M. Serbyn, and R. Vasseur, \textit{{Distinguishing localization from chaos: Challenges in finite-size systems}}, \href{https://doi.org/10.1016/j.aop.2021.168415
}{ Ann. Phys. {\bfseries 427}, 168415 (2021)}, [\href{ 	
https://doi.org/10.48550/arXiv.1911.04501}{{\ttfamily  arXiv:1911.04501}}].
%
\bibitem{Sels:2021}
 D. Sels, A. Polkovnikov, \textit{{Dynamical obstruction to localization in a disordered spin chain}}, \href{https://doi.org/10.1103/PhysRevE.104.054105}{ Phys. Rev. E. {\bfseries 104}, 054105 (2021)}, [\href{ https://doi.org/10.48550/arXiv.2009.04501}{{\ttfamily  arXiv:2009.04501}}].
%
\bibitem{Roeck:2017}
 W. De Roeck, and F. Huveneers, \textit{{Stability and instability towards delocalization in many-body localization systems}}, \href{https://doi.org/10.1103/PhysRevB.95.155129}{ Phys. Rev. B. {\bfseries 95}, 155129 (2017)}, [\href{https://doi.org/10.48550/arXiv.1608.01815}{{\ttfamily arXiv:1608.01815 }}].
\end{thebibliography}

\begin{thebibliography}{10}

\bibitem{Guhr:1998}
T. Guhr,  A. M{\"u}ller-Groeling, and H. A. Weidenm{\"u}ller, \textit{{Random-matrix theories in quantum physics: common concepts}}, \href{https://doi.org/10.1016/S0370-1573(97)00088-4}{Phys. Rep. {\bfseries 299}, 189--425 (1998)}, [\href{https://doi.org/10.48550/arXiv.cond-mat/9707301}{{\ttfamily arXiv:cond-mat/9707301}}].
%
\bibitem{Forrester:2003}
P. J. Forrester, N. C. Snaith, and  J. J. M. Verbaarschot, \textit{{Developments in random matrix theory}}, \href{https://doi.org/10.1088/0305-4470/36/12/201}{ J. Phys. A: Math. Theor. {\bfseries 36}, R1 (2003)}, [\href{ https://doi.org/10.48550/arXiv.cond-mat/0303207}{{\ttfamily arXiv:cond-mat/0303207 }}].
%
%
%
\bibitem{Livan:2018}
G. Livan, M. Novaes, and P. Vivo, \textit{{Introduction to random matrices theory and practice}}, \href{}{Monograph Award., 63 (2018)}, [\href{ https://doi.org/10.48550/arXiv.1712.07903}{{\ttfamily arXiv:1712.07903}}].
%
\bibitem{Atas:2018}
Y. Y. Atas, E. Bogomolny, O. Giraud, and G. Roux, \textit{{Distribution of the ratio of consecutive level spacings in random matrix ensembles}}, \href{ 	
	https://doi.org/10.1103/PhysRevLett.110.084101}{ Phys. Rev. Lett. {\bfseries 110}, 084101 (2018)}, [\href{ 	
	https://doi.org/10.48550/arXiv.1212.5611}{{\ttfamily arXiv:1212.5611}}]. 
%
\bibitem{Miszczak:2012}
J. A. Miszczak, \textit{{Generating and using truly random quantum states in Mathematica}}, \href{https://doi.org/10.1016/j.cpc.2011.08.002}{Comput. Phys. Commun. {\bfseries 183}, 118-124 (2012)}, [\href{https://arxiv.org/pdf/1102.4598.pdf}{{\ttfamily arXiv:1102.4598v2}}].
%
\bibitem{S:Grover:2018}
J. R. Garrison and T. Grover, \textit{{Does a Single Eigenstate Encode the Full Hamiltonian?}}, \href{https://doi.org/10.1103/PhysRevX.8.021026}{Phys. Rev. X. {\bfseries 8}, 021026 (2018)}[\href{https://doi.org/10.48550/arXiv.1503.00729}{{\ttfamily arXiv:1503.00729}}].
%
\bibitem{S:Dymarsky:2016ntg}
A.~Dymarsky, N.~Lashkari and H.~Liu, \textit{{Subsystem eigenstate
  thermalization hypothesis}},
  \href{http://dx.doi.org/10.1103/PhysRevE.97.012140}{Phys. Rev. E {\bfseries
  97}, 012140 (2018)}, [\href{https://arxiv.org/abs/1611.08764}{{\ttfamily
  arXiv:1611.08764}}].
%
  \bibitem{Rigol:2019}
T. LeBlond, K. Mallayya, L. Vidmar, and M. Rigol, \textit{{Entanglement and matrix elements of observables in interacting integrable systems}, Phys. Rev. E. {\bfseries 100}, 062134 (2019)}.

\end{thebibliography}
\end{document}